\documentclass[reprint,aps,prb,amsmath,amssymb]{revtex4-2}

\usepackage[]{graphicx}
\usepackage{times}
\usepackage{bm}
\usepackage[ulem=normalem]{changes}
\usepackage[normalem]{ulem}		
\usepackage{color}
\usepackage{xr} 				
\usepackage{cancel}
\usepackage{hyperref}
\usepackage{siunitx}
\usepackage{babel}
\usepackage{subcaption}
\usepackage{mathtools}
\usepackage{dsfont}
\usepackage{tikz}
\usetikzlibrary{decorations.pathmorphing}
\usetikzlibrary{arrows.meta}
\usetikzlibrary{patterns}
\usetikzlibrary{calc,positioning}

\definechangesauthor[color=blue]{SM}

\usepackage{cleveref}
\usepackage{braket}
\DeclareUnicodeCharacter{2212}{\ensuremath{-}}
\DeclareUnicodeCharacter{00A0}{~}

\begin{document}

\title{Theory of x-ray scattering from optically pumped excitons in atomically thin semiconductors}

\author{Joris Sturm$^{1}$,  Andrei Benediktovitch$^{2}$, Nina Rohringer$^{2,3}$ and Andreas Knorr$^{1}$}
\affiliation{$^1$ Technische Universität Berlin, Institut für Physik und Astronomie, Nichtlineare Optik und Quantenelektronik, 10623 Berlin, Germany}
\affiliation{$^2$ Deutsches Elektronen-Synchrotron DESY, Hamburg 22607, Germany}
\affiliation{$^3$ Department of Physics, Universität Hamburg, Hamburg 22607, Germany}

\begin{abstract}
We propose a framework to explore the internal charge distribution of mesoscopic quasiparticles by inelastic x-ray scattering, while also accounting for the conventional scattering from electrons. Specifically, we investigate a new contribution of intrinsic and optically pumped excitons (bound electron-hole pairs) to the x-ray scattering spectrum of transition metal dichalcogenides (TMDCs). The optical excitation leads to the creation of Wannier exciton populations, adding new quasi-elastic processes beyond the conventional electronic features to the x-ray scattering spectra. Differential spectra (with and without optical pumping) can be used to isolate and identify the internal charge distribution of the optically pumped excitons in the scattering response, potentially offering insights into many-body interactions and quasi-particle dynamics in 2D systems.
\end{abstract}

\maketitle

\section{Introduction}
X-ray scattering is a prime technique to determine atomic and electronic structure in the solid state. The differential scattering cross section of coherent elastic x-ray scattering is linked to the one-particle electron density of the underlying system and is the key technique to ground-state atomic and electronic structure determination in the realm of crystallography. In order to understand a material’s electronic rearrangement due to optical excitations or the inclusion of charged impurities – electronic screening effects – more intricate experimental techniques have been devised that give access to higher-order, nonlocal electron density-density response functions. One of the prime methods is inelastic x-ray scattering (IXS)\cite{schulke_electron_2007} with photon energies far from any x-ray absorption edges: In the impulsive, nonresonant x-ray scattering limit the momentum and energy resolved scattered photon intensity (the dynamic structure factor) encodes the electronic 2-body density-density response function which gives both structural and spectroscopic information of the manifold of excited electronic states. IXS has been used to study momentum-resolved excitation spectra of solids and is a sensitive spectroscopic tool to map out the single- and collective electronic excitations, ranging from chemically sensitive analysis of core \cite{soininen_scheme_2001,rueff_inelastic_2010,sternemann_near-edge_2007} and valence-electronic excitations \cite{feng_exciton_2008, dalecky_non-resonant_2024} to the study of collective excitations such as plasmons \cite{reed_effective_2010} and magnons \cite{hill_observation_2008}. To some extent, also lattice degrees of freedom, i.e. phonon-modes can be measured in high-resolution setups \cite{baron_introduction_2020, krisch_inelastic_2007}. Due to its sensitivity to measure collective and many-body excitations, energy - and momentum resolved IXS constitutes a critical experimental technique to benchmark electronic structure models for dynamic electron screening, i.e. exchange and correlation kernels and many-body effects that can be captured only with methods beyond density functional theory \cite{reshetnyak_excitons_2019, larson_nonresonant_2007, haverkort_nonresonant_2007, shirley__1999, benedict_optical_1998}. \\
In this contribution we theoretically explore IXS on optically excited semiconductors, a process that leads to x-ray optical wavemixing \cite{ornelas-skarin_second-order_2026, glover_x-ray_2012} and hence gives access to fundamentally different electronic response functions from IXS on systems in the electronic ground state \cite{boemer_towards_2021, popova-gorelova_atomic-scale_2024, krebs_theory_2021}. In particular, we focus on imaging the internal spatial charge distributions of optically pumped composite mesoscopic excitonic quasiparticles. 
Previously, IXS has successfully been applied to image excitonic charge response functions \cite{abbamonte_dynamical_2008} without optical excitations of excitons. Within the realm of causality-restraint phase-problem of Fourier transformation, the locally averaged electronic density-density response function can be obtained with a resolution on the sub-fs temporal and Angström length scale \cite{abbamonte_dynamical_2008}. Here, we theoretically study IXS on the steady-state limit of optically excited excitons and express the electronic response function in terms of excitonic wave functions and their internal charge-density distribution. Thereby, we distill the structural information that can be obtained beyond IXS of the unpumped system. We focus on excitons in semiconductors as an example for mesoscopic excitations extended over many atomic elementary cells. We illustrate our theoretical results by modeling Wannier excitons occurring in two-dimensional (2D) materials, such as transition metal dichalcogenides (TMDCs). These materials have emerged as a vibrant platform for exploring mesoscopic quantum phenomena, such as excitonic phase transitions \cite{liu_excitonic_2023, katzer_exciton-phonon_2023} and spatial coherence \cite{troue_extended_2023}. The strong light-matter interaction\cite{splendiani_emerging_2010,liu_strong_2015,schneider_two-dimensional_2018} and large exciton binding energies \cite{ramasubramaniam_large_2012,chernikov_exciton_2014,steinhoff_efficient_2015,hill_observation_2015}, which are a consequence of the reduced screening in two dimensions, make TMDCs interesting materials to study excitons without the interference of free particle electron-hole interactions. 
\\
The investigated situation is illustrated in Fig \ref{fig:sketch_x-ray_scattering}:
\begin{figure}[ht!]
    \centering
\includegraphics[width=0.99\linewidth]{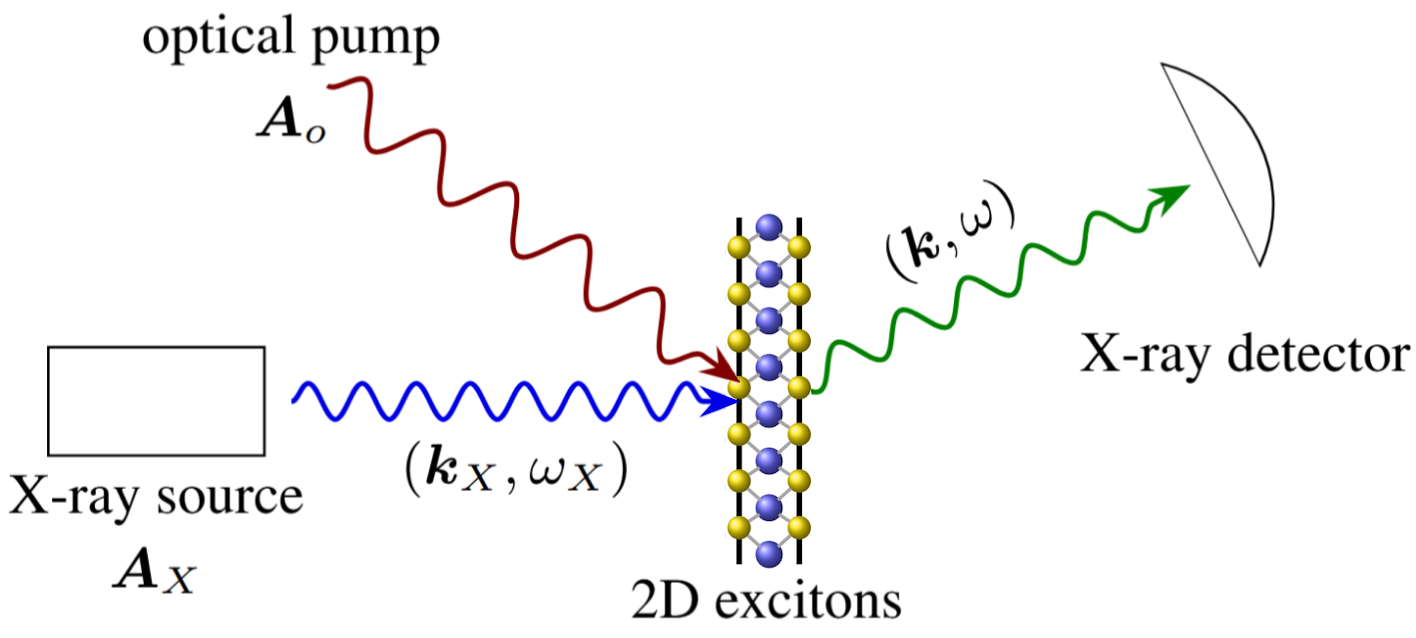}
    \caption{x-ray scattering geometry: A free-standing atomically thin semiconductor is optically excited by a laser with vector potential $\boldsymbol{A}_{o}$ at a frequency resonant with the band gap of the material, creating excitons.  The incoming x-ray field with vector potential  $\boldsymbol{A}_{X}$ of wavevector $\boldsymbol{k}_{X}$ and frequency $\omega_{X}$ is scattered by the excited electronic system. The scattered field is frequcency resolved at different wavevectors $\boldsymbol{k}$, giving the spectrum $S(\Delta\boldsymbol{q},\omega)$ at a particular momentum transfer $\Delta\boldsymbol{q}=\boldsymbol{k}_{X}-\boldsymbol{k}$.}
    \label{fig:sketch_x-ray_scattering}
\end{figure}
 An incoming x-ray field $\boldsymbol{A}_{X}$ with frequency $\omega_{X}$ and wavevector $\boldsymbol{k}_{X}$ is scattered by an excitonic population in the atomically thin semiconductor, excited by an optical field $\boldsymbol{A}_{o}$. The scattered x-ray field with frequency $\omega$ and wavevector $\boldsymbol{k}$ can be subsequently detected by an x-ray detector.
\section{Theoretical framework}
\subsection{Time and spectrally resolved intensity}
To study signatures of optically excited semiconductor excitons in x-ray scattering, we calculate the steady-state limit of the spectrally resolved x-ray yield $S(\boldsymbol{R};\omega)$ \cite{vogel_quantum_2006} at frequency $\omega$ and observation point $\boldsymbol{R}$, cp. Fig. \ref{fig:sketch_x-ray_scattering}:
\begin{equation}
S(\boldsymbol{R};\omega) = \frac{1}{\Gamma}\lim_{t\to\infty} I(t;\boldsymbol{R}, \omega), 
\end{equation}
where the spectrally filtered intensity $I(t;\boldsymbol{R}, \omega) $ reads
\begin{equation}
\label{eq:total_spectrum_first_expression}
I(t;\boldsymbol{R}, \omega) \!=\!\! \int\! dt' \! \int \! dt'' T^{*}_{\Gamma}(t\!-\!t') T_{\Gamma}(t\!-\!t'')   G(\boldsymbol{R};t',t'').
\end{equation}
$T_{\Gamma}(t)$ represents the response function of the spectral filter
\begin{equation}
    T_{\Gamma}(t) = \Gamma \, \theta(t) e^{\!-(i\omega + \Gamma) t}
\end{equation}
with central frequency $\omega$ and spectral width $\Gamma$. Furthermore, in Eq. \ref{eq:total_spectrum_first_expression}
$G(\boldsymbol{R};t',t'')$ is the time-ordered correlation function of the x-ray electric-field operators $\boldsymbol{E}^{(-)}_{X},\boldsymbol{E}^{(+)}_{X}$
\begin{equation}
\label{eq:definition_correlation_function_electric_fields}
    G(\boldsymbol{R};t',t'') = \braket{\boldsymbol{E}^{(-)}_{X}(\boldsymbol{R},t') \boldsymbol{E}^{(+)}_{X}(\boldsymbol{R},t'')}
\end{equation}
where $\boldsymbol{E}^{(-)}_{X}(\boldsymbol{R},t) = \bigl(\boldsymbol{E}^{(+)}_{X}(\boldsymbol{R},t)\bigr)^{\dagger}$ and  $\boldsymbol{E}^{(-)}(\boldsymbol{R},t)$ represents the negative frequency part of the emitted field\cite{vogel_quantum_2006}.\\
In the steady-state limit Eq. \ref{eq:total_spectrum_first_expression} takes the form\cite{vogel_quantum_2006}
\begin{equation}
\label{eq:spectrum_first_definition}
S(\boldsymbol{R};\omega) = \!\int_{0}^{\infty}\! d\tau e^{\!-i\omega  \tau} e^{ -\Gamma \tau}  G(\boldsymbol{R};\tau) +c.c.
\end{equation}
where the time-ordered correlation function $G(\boldsymbol{R};\tau) = \lim_{t\to\infty} G(\boldsymbol{R};t\!+\! \tau,t)$ depends only on the time difference $\tau= t' -t''$. In the following section, we describe the light-matter Hamiltonian that determines $G(\boldsymbol{R};\tau)$.

\subsection{Hamiltonians}
The field-correlation function $G(\boldsymbol{R};\tau)$ of our model is determined by a Hamiltonian $H$ consisting of four contributions: $H=H_{0}+H_{\text{coul}}+H^{X}_{l-m}+H^{o}_{l-m}$. The Hamiltonian $H_{0}$ describes the Bloch electrons in the semiconductor bandstructure:
\begin{equation}
\label{eq:definition_free_Hamiltonian}
    \begin{aligned}
       H_0 =& \sum_{\lambda,\boldsymbol{k_1}} \epsilon^{\lambda}_{\boldsymbol{k_1}} a^{\dagger \lambda}_{ \boldsymbol{k_1}} a^{\lambda}_{\boldsymbol{k_1}}.
    \end{aligned}
\end{equation}
Here, $\lambda$ is the bandindex, $\boldsymbol{k_1}$ the wavevector, $\epsilon^{\lambda}_{\boldsymbol{k}}$ the energy of a Bloch electron and $a^{\lambda}_{ \boldsymbol{k}}$ ($a^{\dagger\lambda}_{ \boldsymbol{k}}$) denote the corresponding fermionic annihilation (creation) operator. The second contribution, $H_{\text{coul}}$, represents the Coulomb interaction between the Bloch electrons:
\begin{equation}
\label{eq:definition_Coulomb_Hamiltonian}
    \begin{aligned}
    H_{\text{coul}} =& \sum_{\substack{\lambda,\lambda' \\ \boldsymbol{k_1}, \boldsymbol{k_2}, \boldsymbol{q}}} \frac{V_{\boldsymbol{q}}}{2} a^{\dagger\lambda}_{\boldsymbol{k_1}+\boldsymbol{q}}a^{\dagger \lambda'}_{\boldsymbol{k_2}-\boldsymbol{q}} a^{\lambda'}_{\boldsymbol{k_2}} a^{\lambda}_{\boldsymbol{k_1}}
    \end{aligned}
\end{equation}

and is responsible for the formation of excitons. Here, $V_{\boldsymbol{q}}$ is the Fourier transform of the screened Keldysh potential\cite{steinhoff_influence_2014,meckbach_influence_2018, trolle_model_2017,katsch_theory_2018}. Although scattering and diffraction experiments of TMDCs are often performed on substrates\cite{vegso_wide-angle_2022,shaji_orientation_2021}, it has been shown that free-standing samples of TMDCs can be produced by liquid exfoliation \cite{coleman_two-dimensional_2011}. For simplicity, we therefore assume a free-standing TMDC monolayer with the screened 2D Coulomb potential: 
\begin{equation}
    \label{eq:Keldysh_potential}
    \begin{aligned}
        V_{\boldsymbol{q}} =& \frac{e^2}{2\epsilon_{0} A |\boldsymbol{q}| \epsilon(\boldsymbol{q})},
            \end{aligned}
\end{equation}
with the dielectric function $ \epsilon(\boldsymbol{q})$ given by
\begin{equation}
    \label{eq:Keldysh_potential_dielectric_function}
        \begin{aligned}
        \epsilon(\boldsymbol{q}) =& \epsilon_{\text{TMDC}} \frac{1- \tilde{\epsilon}e^{\!-h_{\text{TMDC}}|\boldsymbol{q}|} }{1 + \tilde{\epsilon} e^{\!-h_{\text{TMDC}}|\boldsymbol{q}|} } 
    \end{aligned}
\end{equation}
where $e$ is the elementary charge, $A$ is the area of the monolayer, $\epsilon_{\text{TMDC}} $ is the in-plane dielectric constant of the TMDC, $\tilde{\epsilon}=\frac{\epsilon_{\text{TMDC}} -1}{\epsilon_{\text{TMDC}} +1 }$ and $h_{\text{TMDC}}$ is the thickness of the TMDC.\cite{florian_dielectric_2018}
\\
The Hamiltonians $H^{X}_{l-m}$ and $H^{o}_{l-m}$ describe the light-matter interactions, separated into the interaction of x-rays (Eq. \ref{eq:definition_Hamiltonian_x-ray}) and the optical-pump field (Eq. \ref{eq:definition_Hamiltonian_optical_interaction}) with the Bloch electrons:
\begin{equation}
\label{eq:definition_Hamiltonian_x-ray}
    \begin{aligned}
        H^{X}_{l-m}= \frac{e^2}{2m}\sum_{\substack{\lambda,\lambda' \\ \boldsymbol{k_1},\boldsymbol{k_2}}}  \bra{ \Psi^{\lambda}_{\boldsymbol{k_1} }} \boldsymbol{A}_{X}^2(\boldsymbol{r},t)  \ket{ \Psi^{\lambda'}_{\boldsymbol{k_2}}} a^{\dagger \lambda}_{\boldsymbol{k_1}} a^{\lambda'}_{\boldsymbol{k_2}},
    \end{aligned}
\end{equation}
\begin{equation}
\label{eq:definition_Hamiltonian_optical_interaction}
    \begin{aligned}
        H^{o}_{l-m}=& -\!\frac{i\hbar e}{m}\!\sum_{\substack{\lambda,\lambda' \\ \boldsymbol{k_1},\boldsymbol{k_2}}} \hspace{-0.1cm} \boldsymbol{A}_{o}(t) \!\cdot \! \bra{ \Psi^{\lambda}_{\boldsymbol{k_1} }}  \boldsymbol{\nabla}_{\boldsymbol{r}}  
 \ket{ \Psi^{\lambda'}_{\boldsymbol{k_2}}} a^{\dagger \lambda}_{\boldsymbol{k_1}} a^{\lambda'}_{\boldsymbol{k_2}}
    \end{aligned}
\end{equation}
where $e$ is the elementary charge, $m$ is the free electron mass and $ \Psi^{\lambda}_{\boldsymbol{k_1}}$ are the electron Bloch wavefunctions in band $\lambda$ and wavevector $\boldsymbol{k_1}$.
Supposing x-ray frequencies far above any excitation edges of the material, we neglect the $\boldsymbol{p}\cdot\boldsymbol{A}_{X}$ contribution of the x-ray - matter coupling Hamiltonian $H^{X}_{l-m}$ and include only the term proportional to $\boldsymbol{A}_{X}^2$, which determines the scattering contribution in the x-ray range\cite{alsnielsen_elements_2011}. Conversely, the light-matter interaction Hamiltonian $H^{o}_{l-m}$ is treated in dipole approximation\cite{haug_quantum_2009,kira_semiconductor_2011,scully_quantum_1997}: Eq. \ref{eq:definition_Coulomb_Hamiltonian} and Eq. \ref{eq:definition_Hamiltonian_optical_interaction} describe the generation of excitons by the combined action of Coulomb coupled Bloch electrons (Eq. \ref{eq:definition_free_Hamiltonian}, \ref{eq:definition_Coulomb_Hamiltonian}) and resonant optical pumping. The incoming x-ray and optical vector potentials are assumed to be monochromatic :
\begin{equation}
\label{eq:approximation_A_field}
\begin{aligned}
        \boldsymbol{A}_{X}(\boldsymbol{r},t) =& \boldsymbol{A}_{X} \bigl( e^{i(\boldsymbol{k}_{X}\cdot\boldsymbol{r} - \omega_{X}t)} +c.c)
        \\
         \boldsymbol{A}_{\text{o}}(t) =& \boldsymbol{A}_{\text{o}} (e^{\!-i\omega_{\text{o}}t}+e^{i\omega_{\text{o}}t})     
\end{aligned}
\end{equation}
with the amplitudes $\boldsymbol{A}_{X}, \boldsymbol{A}_{\text{o}}$. The wavevector and frequency of the incoming x-ray field are $\boldsymbol{k}_{X}$ and $\omega_{X}$ and the frequency of the optical pump field is $\omega_{o}<<\omega_{X}$. 

\subsection{Time-ordered correlation function}
The electric field of the scattered x-ray radiation in the far-field $\boldsymbol{E}^{(+)}_{X}(\boldsymbol{R},t) = i\omega_{X} \boldsymbol{A}^{\perp(+)}_{X}(\boldsymbol{R},t)$ at detection point $\boldsymbol{R}$ is determined by the transverse electronic currents of the source points $\boldsymbol{r}'$ through\cite{jackson_klassische_2013}:
\begin{equation}
\hspace{-0.2cm}
\label{eq:electric_field_time_domain_j_perp}
\begin{aligned}
    \boldsymbol{E}^{(+)}_{X}(\boldsymbol{R},t) \!  =&\frac{i\omega_{\hspace{-0.05cm}X} \mu_0 }{4\pi R } e^{i \frac{\omega_{x} R}{c}} \hspace{-0.15cm} \int \! d^3r'  \boldsymbol{j}^{\perp(+)}_{X}\!(\boldsymbol{r}',t) e^{\!-i\boldsymbol{k}\cdot\boldsymbol{r}'}
    \end{aligned}
\end{equation}
where $\omega_{X}$ denotes the frequency of the incoming x-ray, $\boldsymbol{k}=\hat{\boldsymbol{e}}_{\boldsymbol{R}} \frac{\omega}{c}$ is the wavevector after scattering -- note that we treat elastic $\omega=\omega_X$ and inelastic scattering terms $\omega\neq\omega_X$ -- and the transverse current density operator $\boldsymbol{j}^{\perp}_{X}$ in the x-ray range is given by
\begin{equation}
\label{eq:definition_j_perp}
    \boldsymbol{j}^{\perp}_{X}(\boldsymbol{r},\omega)  = (1- \boldsymbol{\hat{n}}\boldsymbol{\hat{n}})\cdot \boldsymbol{j}_{X}(\boldsymbol{r},\omega).
\end{equation}
Here, $\boldsymbol{\hat{n}} = \frac{\hat{\boldsymbol{R}}}{|\boldsymbol{R}|}$ is the unit vector pointing in the direction of the detector. The current induced by the x-ray field can be determined by $\boldsymbol{j}_{X}(\boldsymbol{r},t) =  -\frac{\partial \mathcal{H}}{\partial \boldsymbol{A}_{X}(\boldsymbol{r},t)}$:
\begin{equation}
\label{eq:current_density_operator_x-ray}
   \begin{aligned}
      \boldsymbol{j}_{X}(\boldsymbol{r},t)  = -\frac{e^2}{m}\!\sum_{\substack{\lambda,\lambda' \\ \boldsymbol{k_1},\boldsymbol{k_2}}} \Psi^{*\, \lambda}_{\boldsymbol{k_1}}(\boldsymbol{r})  \boldsymbol{A}_{X}(\boldsymbol{r},t) \Psi^{\lambda'}_{\boldsymbol{k_2}}(\boldsymbol{r})  a^{\dagger \lambda}_{\boldsymbol{k_1}} a^{\lambda'}_{\boldsymbol{k_2}} 
   \end{aligned} 
\end{equation}
Eq. \ref{eq:current_density_operator_x-ray} determines the spectrum $S(\Delta\boldsymbol{q},\omega)$ via the time-ordered correlation function $\lim_{t\to\infty}   G(\Delta\boldsymbol{q},t\!+\! \tau,t)$ Eq. \ref{eq:definition_correlation_function_electric_fields} 
    \begin{equation}
\label{eq:definition_of_G_current_density_correlation}
\begin{aligned}    
G(\Delta\boldsymbol{q},t\!+\! \tau,t)=& \left|\boldsymbol{A}^{\! \perp}_{\! X} \omega_{X}\! \dfrac{ r_{0}}{ R }\right|^2 \hspace{-0.1cm} \sum_{\substack{\lambda,\lambda' \\ \nu,\nu' \\ \boldsymbol{k_1},\boldsymbol{k_2} \\ \boldsymbol{k_3},\boldsymbol{k_4} }} \!  s^{\lambda, \lambda',\nu, \nu'}_{\boldsymbol{k_1},\boldsymbol{k_2},\boldsymbol{k_3},\boldsymbol{k_4}}\!(\Delta\boldsymbol{q})  e^{i\omega_{X} \tau}
\\
\times&\braket{ a^{\dagger \lambda}_{\boldsymbol{k_1}}(t\!+\! \tau) a^{\lambda'}_{\boldsymbol{k_2}}(t\!+\! \tau) a^{\dagger \nu}_{\boldsymbol{k_3}}(t)a^{\nu'}_{\boldsymbol{k_4}}(t) },
    \end{aligned}
\end{equation}
where the distance between the target and the detector is denoted by $R$, $r_{0}=\frac{ \mu_0 e^2 }{4\pi m }$ is the Thomson scattering length\cite{alsnielsen_elements_2011} and $\Delta \boldsymbol{q}= \boldsymbol{k}_{X} -\boldsymbol{k}$ is the photon momentum transfer, i.e. the difference of the incoming and scattered wavevector (detection direction) cp. Fig \ref{fig:sketch_x-ray_scattering}, which is the typical quantity observed in x-ray scattering experiments\cite{galambosi_anisotropic_2011,nicolaou_direct_2025}. 
The scattering matrix: $s^{\lambda, \lambda',\nu, \nu'}_{\boldsymbol{k_1},\boldsymbol{k_2},\boldsymbol{k_3},\boldsymbol{k_4}}(\Delta\boldsymbol{q})$, introduced in Eq. \ref{eq:definition_of_G_current_density_correlation} reads:
\begin{equation}
\label{eq:scattering_matrix_with_delta_distributions_and_Bloch_factors}
    \begin{aligned}
       &s^{\lambda, \lambda',\nu, \nu'}_{\boldsymbol{k_1},\boldsymbol{k_2},\boldsymbol{k_3},\boldsymbol{k_4}}(\Delta\boldsymbol{q})
       \\
       =& \delta_{\Delta \boldsymbol{q} + \boldsymbol{k_1},\boldsymbol{k_2}}  \dfrac{1}{A_{UC}} \int_{UC}d^2r' u^{*\, \lambda}_{\boldsymbol{k_1}}(\boldsymbol{r}') u^{\lambda'} _{\boldsymbol{k_2}}(\boldsymbol{r}')
       \\
       \times&\delta_{\Delta \boldsymbol{q} + \boldsymbol{k_4},\boldsymbol{k_3}}\dfrac{1}{A_{UC}} \int_{UC}d^2r'' u^{*\, \nu}_{\boldsymbol{k_3}}(\boldsymbol{r}'') u^{\nu'}_{\boldsymbol{k_4}}(\boldsymbol{r}'').
    \end{aligned}
\end{equation}
It is to be noted that, henceforth, $\boldsymbol{r}$ is a two-dimensional quantity within the plane of the quasi 2D TMDC monolayer.\\
The wavefunction, to calculate Eq. \ref{eq:current_density_operator_x-ray},\ref{eq:definition_of_G_current_density_correlation} is given by $\Psi^{\lambda}_{\boldsymbol{k}}(\boldsymbol{r}) = \frac{1}{\sqrt{A}}e^{i\boldsymbol{k}\cdot\boldsymbol{r}} u^{\lambda}_{\boldsymbol{k}}(\boldsymbol{r}) \xi(z)$. $A_{UC}$ denotes the surface of the unit cell, with the surface of the 2D crystal  $A=N A_{UC} $ where $N$ denotes the number of unit cells. The confinement function $\xi(z)$ is assumed to confine the material as $|\xi(z)|^2=\delta(z)$. Furthermore, we neglected in Eq. \ref{eq:scattering_matrix_with_delta_distributions_and_Bloch_factors} contributions from finite reciprocal lattice vectors and concentrate our analysis to small momentum transfers assuming $\boldsymbol{G}=\boldsymbol{0}$.\\
Eq. \ref{eq:scattering_matrix_with_delta_distributions_and_Bloch_factors} simplifies the time-ordered correlation function Eq. \ref{eq:definition_of_G_current_density_correlation} to the form:
\begin{widetext}
    \begin{equation}
\label{eq:correlation_function_with_use_of_spatial_integrals}
\begin{aligned}
  G(\Delta \boldsymbol{q},t\!+\! \tau,t)
   \! =\! \left|\boldsymbol{A}^{\! \perp}_{\! X} \omega_{X}\! \frac{ r_{0}}{ R }\right|^2\sum_{\substack{\lambda,\lambda'\\ \nu,\nu' \\ \boldsymbol{k_1}, \boldsymbol{k_2} }}&\! \braket{ a^{\dagger \lambda}_{\boldsymbol{k_1}}(t\!+\! \tau) a^{\lambda'}_{\boldsymbol{k_1}\! + \! \Delta \boldsymbol{q}}(t\!+\! \tau) a^{\dagger \nu}_{\boldsymbol{k_2}+\Delta \boldsymbol{q}}(t)a^{\nu'}_{\boldsymbol{k_2}}(t) } e^{i\omega_{X}\! \tau} 
   \\
   \times&\braket{\lambda,\boldsymbol{k_1}|\lambda', \boldsymbol{k_1}\!+\!\Delta \boldsymbol{q}}  \braket{\nu,\boldsymbol{k_2}\!+\!\Delta\boldsymbol{q}|\nu', \boldsymbol{k_2}}
    \end{aligned}
    \end{equation}
\end{widetext}
where we introduced the notation
\begin{equation}
\label{eq:definition_Bloch_overlap}
  \braket{\lambda,\boldsymbol{k_1}|\lambda', \boldsymbol{k_2}} = \frac{1}{A_{UC}} \int_{UC}d^2r' u^{*\, \lambda}_{\boldsymbol{k_1}}(\boldsymbol{r}') u^{\lambda'}_{\boldsymbol{k_2}}(\boldsymbol{r}')
\end{equation}
and the Bloch factors are normalized according to:
\begin{equation}
    \frac{1}{A_{UC}} \int_{UC}d^2r' |u^{\lambda'}_{\boldsymbol{k_1}}(\boldsymbol{r}')|^2 = 1.
\end{equation}
Eq. \ref{eq:correlation_function_with_use_of_spatial_integrals} determines the spectrum\cite{vogel_quantum_2006}:
\begin{equation}
\label{eq:spectrum_final_definition}
S(\Delta \boldsymbol{q},\omega) = \int_{0}^{\infty} d\tau e^{\!-i\omega  \tau} e^{ -\Gamma \tau}  G(\Delta \boldsymbol{q},\tau) +c.c.
\end{equation}
Eq. \ref{eq:definition_of_G_current_density_correlation} and Eq. \ref{eq:correlation_function_with_use_of_spatial_integrals}-\ref{eq:spectrum_final_definition} form the basis to discuss the x-ray scattering signal from optically pumped excitons in a two-dimensional semiconductor. The excitonic features are included in the two-time correlation $\braket{ a^{\dagger \lambda}_{\boldsymbol{k_1}}(t\!+\! \tau) a^{\lambda'}_{\boldsymbol{k_1}\! + \! \Delta \boldsymbol{q}}(t\!+\! \tau) a^{\dagger \nu}_{\boldsymbol{k_2}+\Delta \boldsymbol{q}}(t)a^{\nu'}_{\boldsymbol{k_2}}(t) }$ and will be analyzed in section \ref{sec:Optically excited two-band semiconductor}.
Just for reference, in Appendix \ref{app:electron_gas_limit} we show that the approach presented here includes the benchmark of x-ray scattering by a free electron gas\cite{sturm_dynamic_1993,cardona_light_1983,schulke_electron_2007}. In the subsequent calculations, we assume a perfect spectral filter and therefore $\lim_{\Gamma \to 0}$.

\section{Optically excited two-band semiconductor}
\label{sec:Optically excited two-band semiconductor}
\begin{figure}[ht!] 
    \centering
    \begin{tikzpicture} 
\fill[blue!50] (.,1.05) circle (0.2) node {$e$};
\node[draw,circle,minimum size=.4cm,inner sep=0pt] at (.,1.05) {$e$};
\fill[red!50] (-.4,-2.6) circle (0.2) node {$h$};
\node[draw,circle,minimum size=.4cm,inner sep=0pt] at (-.4,-2.6) {$h$};

\draw[black,line width=0.5mm] (-2.,-2.6) .. controls (-0.7,-2.3) and (-.2,-2.3) .. (1.1,-2.6);
\draw[black,line width=0.5mm] (-1.7,1.5) .. controls (-.95,0.5) and (.05,0.5) .. (.8,1.5);

\node[draw, dotted] (P1) at (-.45,2.) {$P^{\dagger}_{\boldsymbol{k_1},\boldsymbol{k_2}}$};
\node (e_ck) at (-1.7,1.1) {$\epsilon^{c}_{\boldsymbol{k}}$};
\node (e_vk) at (-1.8,-2.3) {$\epsilon^{v}_{\boldsymbol{k}}$};

\draw[-stealth,black,line width=0.5mm] (-2.1,-3.) -- (1.3,-3.) node[below] (COM) {$\mathbf{k}$};

\draw[-stealth,black,line width=0.5mm] (-2.3,-2.8) -- (-2.3,1.7) node[left] {E};

\draw[-latex,gray,line width=1.mm] (1.4,2) -- (2.4,2) node[right,below] (COM) {};

\draw[-,dashed,black,line width=0.3mm] (-.45,-2.4) -- (-.45,.75) node[left,midway] (COM) {$E_{gap}$};

    \begin{scope}[shift = {(-1., 0)}]

\fill[orange!25, rotate=90] (-.15,-5.3) ellipse (0.3 and 0.6);
\draw[black,rotate=90,dashed] (-.15,-5.3) ellipse (0.3 and .6);

\fill[blue!50] (5.6,-.15) circle (0.2) node {$e$};
\node[draw,circle,minimum size=.4cm,inner sep=0pt] (-) at (5.6,-.15) {$e$};

\fill[red!50] (5,-.15) circle (0.2) node {$h$};
\node[draw,circle,minimum size=.4cm,inner sep=0pt] (+) at (5,-.15) {$h$};

\draw[black,line width=0.5mm] (3.75,.1) .. controls (4.25,-.6) and (6.,-.6) .. (6.5,.1);
\draw[black,dashed,line width=0.3mm] (3.75,.85) .. controls (4.25,.1) and (6.,.1) .. (6.5,.85);
\draw[black,dashed,line width=0.3mm] (4.,.8) .. controls (4.35,.3) and (5.9,.3) .. (6.25,.8);

\node (e_Q) at (6.6,-.3) {$\epsilon^{\nu}_{\boldsymbol{Q}}$};
\node[draw,dotted] (P2) at (5.1,2.) {$P^{\dagger\, \nu}_{\boldsymbol{Q}}$};
\draw[-stealth,black,line width=0.5mm] (3.5,-2.4) -- (6.8,-2.4) node[below] (COM) {$\mathbf{Q}$};

\draw[-,dashed,black,line width=0.3mm] (3.5,-2.4) -- (3.5,.75) node[left,midway] (COM) {$E_{gap}$};

\begin{scope}
    \clip (3.5,.75) rectangle (6.7,1.25);
    \fill[pattern=north west lines, pattern color=gray] (3.5,.75) rectangle (6.7,1.25);
\end{scope}
\draw (3.5,.75) -- (6.7,.75);
\node at (5.15,1.) {continuum};
\end{scope}
\end{tikzpicture}
    \caption{Electron-hole picture vs. exciton picture. The energy of the band gap is shown by a vertical dashed line in both pictures. \textbf{Left:} Bandstructure of a two-band semiconductor in a parabolic approximation with dispersion $\epsilon^{c/v}_{\boldsymbol{k}}$ of the conduction/valence band. An electron with momentum $\boldsymbol{k_1}$ is created in the conduction band and a hole with momentum $\boldsymbol{k_2}$ is created in the valence band. This transition can be described by the pair-operator $P^{\dagger}_{\boldsymbol{k_1},\boldsymbol{k_2}}$. \textbf{Right:} A Coulomb-bound electron-hole pair, an exciton, is created with center-of-mass momentum $\boldsymbol{Q}=\boldsymbol{k_1}-\boldsymbol{k_2}$ and quantum number $\nu$. This can be described by the exciton operator $P^{\dagger\, \nu}_{\boldsymbol{Q}}$. The exciton dispersion is given by $\epsilon^{\nu}_{\boldsymbol{Q}}$. Dispersions of other excitonic quantum numbers are indicated as dashed lines. Above the band gap energy the continuum is visualized by the hatched region.}
\label{fig:electron_hole_to_exciton}
\end{figure}
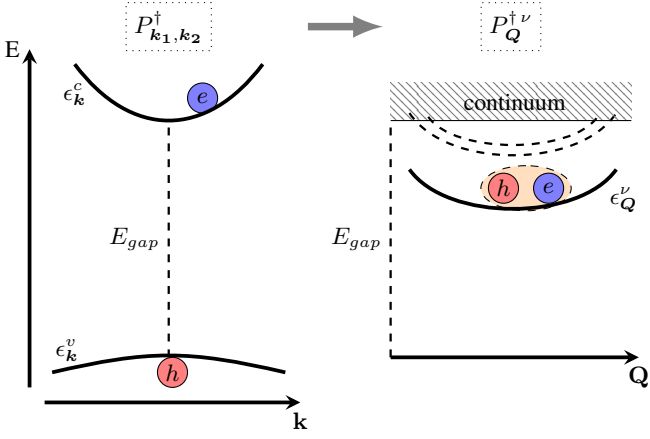

In this section, we study a two-band semiconductor and develop a basic description of excitonic signatures in the x-ray scattering spectrum, Eq. \ref{eq:spectrum_final_definition}.
To include optical excitations of electron-hole pairs illustrated in Fig \ref{fig:electron_hole_to_exciton} we introduce pair-operators $P^{(\dagger)}_{\boldsymbol{k_1},\boldsymbol{k_2}} $ by combining valence-electron annihilation operators $v_{\boldsymbol{k}}$ and conduction-electron creation operators $c^{\dagger}_{\boldsymbol{k}}$ fulfilling:
\begin{equation}
\label{eq:pair_operator_expansion}
    \begin{aligned}
        c^{\dagger}_{\boldsymbol{k_1}} v_{\boldsymbol{k_2}}&=P^{\dagger}_{\boldsymbol{k_1},\boldsymbol{k_2}} 
        \\
        v^{\dagger}_{\boldsymbol{k_2}} c_{\boldsymbol{k_1}} &= P_{\boldsymbol{k_1},\boldsymbol{k_2}}
        \\
        c^{\dagger}_{\boldsymbol{k_1}} c_{\boldsymbol{k_2}} &= \sum_{\boldsymbol{k}} P^{\dagger}_{\boldsymbol{k_1},\boldsymbol{k}} P_{\boldsymbol{k_2},\boldsymbol{k}}
        \\
        v_{\boldsymbol{k_1}} v^{\dagger}_{\boldsymbol{k_2}} &= \sum_{\boldsymbol{k}} P^{\dagger}_{\boldsymbol{k},\boldsymbol{k_1}} P_{\boldsymbol{k},\boldsymbol{k_2}}
    \end{aligned}
\end{equation}
The pair operator $P^{\dagger}_{\boldsymbol{k_1},\boldsymbol{k_2}} $ describes the interband excitation and creates an electron-hole pair, cp. Fig. \ref{fig:electron_hole_to_exciton}, with electron at $\boldsymbol{k_1}$ and hole at $\boldsymbol{k_2}$. The intraband coherence in the conduction band is described by $c^{\dagger}_{\boldsymbol{k_1}} c_{\boldsymbol{k_2}}$ and the associated electron occupation by $c^{\dagger}_{\boldsymbol{k_1}} c_{\boldsymbol{k_1}}$. Analogously, $v_{\boldsymbol{k_1}} v^{\dagger}_{\boldsymbol{k_2}}$ describes the intraband coherence in the valence band and $v_{\boldsymbol{k_1}} v^{\dagger}_{\boldsymbol{k_1}}$ the hole occupation.
Eq. \ref{eq:pair_operator_expansion} is used to expand the correlation function Eq. \ref{eq:correlation_function_with_use_of_spatial_integrals} for a two-band semiconductor by taking the following operator configurations into account:
\begin{widetext}
\begin{equation}
\label{eq:possibilities_electronic_operators}
\braket{ a^{\dagger \lambda}_{\boldsymbol{k_1}}(t\!+\! \tau) a^{\lambda'}_{\boldsymbol{k_2}}(t\!+\! \tau)  a^{\dagger \nu}_{\boldsymbol{k_3}}(t)a^{\nu'}_{\boldsymbol{k_4}}(t) }=\begin{cases}
            \braket{    v^{\dagger}_{\boldsymbol{k_1}} (t\!+\! \tau) c_{\boldsymbol{k_2}} (t\!+\! \tau) c^{\dagger}_{\boldsymbol{k_3}}(t)v_{\boldsymbol{k_4}}(t) }& \\
           \braket{ c^{\dagger}_{\boldsymbol{k_1}}(t\!+\! \tau) v_{\boldsymbol{k_2}}(t\!+\! \tau)  v^{\dagger}_{\boldsymbol{k_3}}(t)c_{\boldsymbol{k_4}}(t) } & \\
          \braket{   v^{\dagger}_{\boldsymbol{k_1}}(t\!+\! \tau) v_{\boldsymbol{k_2}}(t\!+\! \tau)  v^{\dagger}_{\boldsymbol{k_3}}(t)v_{\boldsymbol{k_4}}(t) }& \\
   \braket{ c^{\dagger}_{\boldsymbol{k_1}}(t\!+\! \tau) c_{\boldsymbol{k_2}}(t\!+\! \tau)  c^{\dagger}_{\boldsymbol{k_3}}(t)c_{\boldsymbol{k_4}}(t) }& \\
    \braket{   c^{\dagger}_{\boldsymbol{k_1}}(t\!+\! \tau) c_{\boldsymbol{k_2}}(t\!+\! \tau)  v^{\dagger}_{\boldsymbol{k_3}}(t)v_{\boldsymbol{k_4}}(t) }& \\
     \braket{  v^{\dagger}_{\boldsymbol{k_1}}(t\!+\! \tau) v_{\boldsymbol{k_2}}(t\!+\! \tau)  c^{\dagger}_{\boldsymbol{k_3}}(t)c_{\boldsymbol{k_4}}(t) }& \\
  \end{cases}
\end{equation}
\end{widetext}
Here, only operator combinations including pair-operator densities ($P^{\dagger}P$) to lowest order in the vacuum or in the external optical pumping field\cite{katsch_theory_2018} are taken into account and higher order nonlinear excitonic combinations are neglected. Neglected terms include e.g. $v^{\dagger}vv^{\dagger}c$ which is of third order.\\
Selfconsistently, in this second order approximation we calculate the Heisenberg equations of motion for the pair-operator $P^{\dagger}_{\boldsymbol{k_1},\boldsymbol{k_2}} $\cite{katsch_theory_2018} from the Hamiltonians Eq. \ref{eq:definition_free_Hamiltonian}-\ref{eq:definition_Hamiltonian_optical_interaction}:
\begin{equation}
\begin{aligned}
    \label{eq:time_evolution_of_electron_hole_pair_operator}
    -&i\hbar\frac{d}{dt} P^{\dagger}_{\boldsymbol{k_1},\boldsymbol{k_2}}  
    \\
    &= (\epsilon^{c}_{\boldsymbol{k_1}}-\epsilon^{v}_{\boldsymbol{k_2}} + \sum_{\boldsymbol{q}} V_{\boldsymbol{q}})P^{\dagger}_{\boldsymbol{k_1},\boldsymbol{k_2}}  - \sum_{\boldsymbol{q}} V_{\boldsymbol{q}} P^{\dagger}_{\boldsymbol{k_1}-\boldsymbol{q},\boldsymbol{k_2} - \boldsymbol{q}} 
    \\
    &+ \frac{i\hbar e}{m} \boldsymbol{A}_{o}(t) \cdot \braket{\Psi^{v}_{\boldsymbol{k_2}}| \boldsymbol{\nabla}_{\boldsymbol{r}}|\Psi^{c}_{\boldsymbol{k_1}}} + \mathcal{O}(P^{\dagger}P).
\end{aligned}
\end{equation}
which covers both, the solution without optical field (vacuum) and finite external optical field.
The expression $\mathcal{O}(P^{\dagger}P)$ denotes higher order terms from the pair-operator expansion, which are neglected in the following.
To expand Eq. \ref{eq:time_evolution_of_electron_hole_pair_operator} into an excitonic basis, we introduce relative- and center of mass coordinates $\boldsymbol{Q}=\boldsymbol{k_1} - \boldsymbol{k_2}$, $\boldsymbol{\kappa}=\beta \boldsymbol{k_1}+\alpha\boldsymbol{k_2}$,  where $\alpha = \frac{m_e}{m_e+ m_h}$, $\beta = \frac{m_h}{m_e+m_h}$ and $m_e/m_h$ are the effective masses of electron or hole, respectively. Utilizing this, we can express the pair-operators $P^{(\dagger)}_{\boldsymbol{k_1},\boldsymbol{k_2}} $ in terms of the exciton wave functions together with the exciton operators $P^{(\dagger)\, \nu}_{\boldsymbol{Q}}$ creating excitons in state $\nu$, with the wave function $\varphi^{\nu}_{ \boldsymbol{\kappa}}$ fulfilling:
\begin{equation}
\label{eq:pair-operators_to_exciton_operators}
\begin{aligned}
P^{\dagger}_{\boldsymbol{k_1},\boldsymbol{k_2}} &= \sum_{\nu} \varphi^{*\,\nu}_{ \beta \boldsymbol{k_1}+\alpha\boldsymbol{k_2}} P^{\dagger\, \nu}_{\boldsymbol{k_1} - \boldsymbol{k_2}} \equiv \sum_{\nu} \varphi^{*\,\nu}_{ \boldsymbol{\kappa}} P^{\dagger\, \nu}_{\boldsymbol{Q}}
     \\
P_{\boldsymbol{k_1},\boldsymbol{k_2}} &= \sum_{\nu} \varphi^{\nu}_{ \beta \boldsymbol{k_1}+\alpha\boldsymbol{k_2}} P^{ \nu}_{\boldsymbol{k_1} - \boldsymbol{k_2}} \equiv \sum_{\nu} \varphi^{\nu}_{ \boldsymbol{\kappa}} P^{ \nu}_{\boldsymbol{Q}}
    \end{aligned}
\end{equation}
The excitonic wavefunctions $\varphi^{\nu}_{\boldsymbol{\kappa}} $ in Eq. \ref{eq:pair-operators_to_exciton_operators} are solutions of the Wannier equation \cite{kira_semiconductor_2011,haug_quantum_2009}, identified by inspecting Eq. \ref{eq:time_evolution_of_electron_hole_pair_operator}:
\begin{equation}
\label{eq:wannier_equation}
    \frac{\hbar^2 \kappa^2}{2 \mu} \varphi^{\nu}_{\boldsymbol{\kappa}}  - \sum_{\boldsymbol{q}} V_{\boldsymbol{q}} \varphi^{\nu}_{\boldsymbol{\kappa}+\boldsymbol{q}} = \Delta \epsilon^{\nu} \varphi^{\nu}_{\boldsymbol{\kappa}}
\end{equation}
where $\frac{1}{\mu} = \frac{1}{m_e} + \frac{1}{m_h} $ and $ \Delta \epsilon^{\nu}$ is the exciton binding energy. 
The wavefunctions are normalized according to \cite{haug_quantum_2009, kira_semiconductor_2011}:
\begin{equation}
    \begin{aligned}
        &\sum_{\boldsymbol{\kappa}} \varphi^{*\, \nu}_{\boldsymbol{\kappa}} \varphi^{\mu}_{\boldsymbol{\kappa}} \hspace{0.12cm}= \delta^{\nu,\mu}
        \\
        &\sum_{\nu} \varphi^{*\, \nu}_{\boldsymbol{\kappa}_1} \varphi^{\nu}_{\boldsymbol{\kappa}_2} = \delta_{\boldsymbol{\kappa}_1,\boldsymbol{\kappa}_2}
    \end{aligned}
\end{equation}
Using Eq. \ref{eq:time_evolution_of_electron_hole_pair_operator}-\ref{eq:wannier_equation}, we obtain an equation for $P^{^\dagger \, \nu}_{\boldsymbol{Q}}(t)$:
\begin{equation}
\label{eq:HEOM_exciton}
    \begin{aligned}
        \frac{d}{dt} P^{\dagger\, \nu}_{\boldsymbol{Q}}(t)  &=i\Omega^{\nu}_{\boldsymbol{Q}} P^{\dagger \, \nu}_{\boldsymbol{Q}}(t) - i \delta_{\boldsymbol{Q},\boldsymbol{0}} \boldsymbol{\Omega}^{\nu}_{v,c} \cdot \boldsymbol{E}^{(-)}_{o}(t).
    \end{aligned}
\end{equation}
Here, $\hbar \Omega^{\nu}_{\boldsymbol{Q}}=\epsilon^{\nu}_{x}+ \frac{\hbar^2 \boldsymbol{Q}^2}{2M}$ is the energy dispersion of the exciton with $M=m_{e}+m_{h}$ and $\epsilon^{\nu}_{x} - E_{gap} = \Delta \epsilon^{\nu}$, where $E_{gap}$ denotes the energy of the band gap. 
In Eq. \ref{eq:HEOM_exciton} we rewrote the coupling to the optical pump field in dipole approximation via the relation of the optical matrix element and the momentum matrix element $\bra{ \Psi^{v}} \boldsymbol{p} \ket{ \Psi^{c}} \approx -i\frac{\omega_{o} m}{q} \, \boldsymbol{d}_{v,c}$. Since the optical field is taken as spatially constant, only vanishing momentum transfer $\boldsymbol{Q}=\boldsymbol{0}$ processes contribute to the coupling with the optical pump field.
\\
In a rotating wave approximation\cite{vogel_quantum_2006,haug_quantum_2009}, the optical field reads $\boldsymbol{E}^{\pm}_{o} = \mp i\omega_{o} \boldsymbol{A}^{\pm}_{o} $.
In addition, we used the definition $\boldsymbol{\Omega}^{\nu}_{v,c} = \frac{ \boldsymbol{d}_{v,c} \sum_{\boldsymbol{\kappa}} \varphi^{\nu}_{\boldsymbol{\kappa}}}{\hbar}$ with the optical matrix element $\boldsymbol{d}_{v,c}$ and the relation $\boldsymbol{\Omega}^{\nu}_{v,c} = (\boldsymbol{\Omega}^{\nu}_{c,v} )^* $. 

Eq. \ref{eq:HEOM_exciton} describes the equations of motion for the exciton in a closed system. However, dissipative processes such as radiative dephasing or exciton-phonon interaction are necessary to establish the steady state. 
The corresponding dephasing rates $\gamma^{\nu}_{\boldsymbol{Q}}$ are well known and can be calculated from a microscopic basis\cite{selig_excitonic_2016}. 
The radiative dephasing rate is given by:
\begin{equation}
    \gamma^{\nu}_{\text{rad}} = \frac{\omega_{o} \mu_{0} c}{\hbar A n}\biggl|\boldsymbol{d}_{c,v} \sum_{\boldsymbol{\kappa}}\varphi^{\nu}_{\boldsymbol{\kappa}} \biggr|^2
\end{equation}
with the vacuum permeability $\mu_{0}$ and the velocity of light in the substrate material given by $\frac{c}{n}$.\\
The non-radiative dephasing from exciton-phonon interaction is given by\cite{selig_excitonic_2016}:
\begin{equation}
    \gamma^{\nu,\alpha}_{\boldsymbol{Q}} = \frac{\pi}{\hbar^2} \sum_{\boldsymbol{q},\mu}G^{\nu,\mu, \alpha}_{\boldsymbol{Q},\boldsymbol{Q}+\boldsymbol{q}}
\end{equation}
where $\nu$ and $\mu$ are excitonic states, $\alpha$ is the phonon-mode and $G^{\nu,\mu,\alpha}_{\boldsymbol{Q},\boldsymbol{Q}+\boldsymbol{q}}$ is given by:
\begin{equation}
    \begin{aligned}
  G^{\nu,\mu, \alpha}_{\boldsymbol{Q},\boldsymbol{Q}+\boldsymbol{q}}\! =\!  \sum_{\pm, \lambda}\! g^{\nu,\lambda,\alpha}_{\boldsymbol{q}}\! g^{\lambda,\mu,\alpha}_{-\boldsymbol{q}}\! \bigl( \frac{1}{2}\! \pm\! \frac{1}{2}\! +\! n^{\alpha}_{\boldsymbol{q}}\bigr) \delta\!\left( \Omega^{\lambda}_{\boldsymbol{Q}+\boldsymbol{q}}\! -\!  \Omega^{\mu}_{\boldsymbol{Q}}\! \pm\! \omega^{\alpha}_{\pm \boldsymbol{q}}\right) 
    \end{aligned}
\end{equation}
with phonon energy $\hbar \omega^{\alpha}_{\boldsymbol{q}}$ and phonon occupation $n^{\alpha}_{\boldsymbol{q}}$. Included are phonon emission (+) and absorption (-) processes. The exciton-phonon coupling $g^{\nu,\lambda,\alpha}_{\boldsymbol{q}} $ is given by:
\begin{equation}
    g^{\nu,\lambda,\alpha}_{\boldsymbol{q}} = \sum_{\boldsymbol{q'}} \left( \varphi^{*\, \nu}_{\boldsymbol{q'}} g^{c, \alpha}_{\boldsymbol{q}} \varphi^{\lambda}_{\boldsymbol{q'}-\beta \boldsymbol{q}} - \varphi^{*\, \nu}_{\boldsymbol{q'}} g^{v, \alpha}_{\boldsymbol{q}} \varphi^{\lambda}_{\boldsymbol{q'}+\alpha \boldsymbol{q}} \right)
\end{equation}
This coupling element is dependent on the electron-phonon matrix elements $ g^{c, \alpha}_{\boldsymbol{q}}$ and $ g^{v, \alpha}_{\boldsymbol{q}}$ for the conduction and valence band and the overlap of excitonic wavefunctions \cite{selig_excitonic_2016}.\\
We describe the stationary state of the excitonic excitation in Eq. \ref{eq:HEOM_exciton} in the optical range by the corresponding dephasing $\gamma^{\nu}_{\boldsymbol{Q}}/2$ and employ a Heisenberg-Langevin approach \cite{gardiner_quantum_2004,scully_quantum_1997} to generalize Eq. \ref{eq:HEOM_exciton} by including dissipation and fluctuation forces:
\begin{equation}
\label{eq:heisenberg-langevin_equation_exciton}
    \begin{aligned}
        \frac{d}{dt} P^{\dagger\, \nu}_{\boldsymbol{Q}}(t)  &=\bigl(i\Omega^{\nu}_{\boldsymbol{Q}}-\frac{\gamma^{\nu}_{\boldsymbol{Q}}}{2}\bigr) P^{\dagger \, \nu}_{\boldsymbol{Q}}(t) 
        \\
        &- i \delta_{\boldsymbol{Q},\boldsymbol{0}} \boldsymbol{\Omega}^{\nu}_{v,c} \cdot \boldsymbol{E}^{(-)}_{o}(t) +F^{\dagger\, \nu}_{\boldsymbol{Q}}(t).
    \end{aligned}
\end{equation}
Using Eq. \ref{eq:heisenberg-langevin_equation_exciton} one can generalize the commutation relation to
\begin{equation}
\label{eq:generalized_commutator}
    \begin{aligned}
    \relax
[P^{\mu}_{\boldsymbol{Q'}}(t),P^{\dagger \, \nu}_{\boldsymbol{Q}}(t\!+\! \tau) ]  
     =\delta^{\nu,\mu}_{\boldsymbol{Q},\boldsymbol{Q'}}e^{i\Omega^{\nu}_{\boldsymbol{Q}}\tau} e^{\!-\frac{\gamma^{\nu}_{\boldsymbol{Q}}}{2}\tau}  
    \end{aligned}
\end{equation}
The dissipative term results in commuting operators for large time differences $\tau \to \infty$. Also, in the limit $\tau \to 0$ the bosonic commutation relation for the excitons is recovered. For the numerical calculations we assume in the following that the dephasing is not dependent on wavevector $\boldsymbol{Q}$ nor the excitonic state $\nu$, so $\gamma^{\nu}_{\boldsymbol{Q}}\approx \gamma$.

\section{Results}
\subsection{Optical pump off: Intrinsic excitonic spectrum}
In this section, we discuss x-ray scattering from the intrinsic semiconductor (filled valence band; empty conduction band) i.e. without any optical pumping.
In this case, the contributing processes are the first (pair-antipair vacuum fluctuations) and third line (valence band electrons) in Eq. \ref{eq:possibilities_electronic_operators}. To contribute, all other combinations in Eq. \ref{eq:possibilities_electronic_operators} require direct optical excitation discussed in sec. \ref{sec:optically_pumped_excitonic_spectrum}. Exemplarily we calculate the contribution from the first line and rewrite the electronic operators to excitonic operators as follows: 
\begin{equation}
    \begin{aligned}
        &\braket{v^{\dagger}_{\boldsymbol{k_1}}(t\!+\! \tau) c_{\boldsymbol{k_1}+\Delta\boldsymbol{q}}(t\!+\! \tau)  c^{\dagger}_{\boldsymbol{k_2}+\Delta\boldsymbol{q}}(t) v_{\boldsymbol{k_2}}(t)}
        \\
        =&\sum_{\nu,\nu'} \varphi^{\nu}_{\boldsymbol{k_1}+ \beta \Delta\boldsymbol{q}} \varphi^{*\,\nu'}_{ \boldsymbol{k_2}+\beta \Delta\boldsymbol{q}} \braket{ P^{ \nu}_{\Delta\boldsymbol{q}}(t\!+\! \tau)P^{\dagger\, \nu'}_{\! \Delta\boldsymbol{q}}(t)}
    \end{aligned}
\end{equation}
Applying normal order by using the commutator in Eq. \ref{eq:generalized_commutator} results in two contributions:
\begin{equation}
\label{eq:first_contribution_of_electronic_possibilities_to_excitons}
    \begin{aligned}
&\sum_{\nu,\nu'} \varphi^{\nu}_{\boldsymbol{k_1}+ \beta \Delta\boldsymbol{q}} \varphi^{*\,\nu'}_{ \boldsymbol{k_2}+\beta \Delta\boldsymbol{q}} \braket{ P^{ \nu}_{\Delta\boldsymbol{q}}(t\!+\! \tau)P^{\dagger\, \nu'}_{\! \Delta\boldsymbol{q}}(t)}
\\
=&\sum_{\nu,\nu'}\varphi^{*\,\nu'}_{ \boldsymbol{k_2}+\beta \Delta\boldsymbol{q}}  \varphi^{\nu}_{\boldsymbol{k_1}+ \beta \Delta\boldsymbol{q}} \braket{ P^{\dagger\, \nu'}_{\! \Delta\boldsymbol{q}}(t) P^{ \nu}_{\Delta\boldsymbol{q}}(t\!+\! \tau)}
\\
+&\sum_{\nu} \varphi^{\nu}_{\boldsymbol{k_1}+ \beta \Delta\boldsymbol{q}} \varphi^{*\,\nu}_{ \boldsymbol{k_2}+\beta \Delta\boldsymbol{q}}  e^{\!-i\Omega^{\nu}_{\Delta\boldsymbol{q}} \tau} e^{\!-\frac{\gamma^{\nu}_{\Delta\boldsymbol{q}}}{2}\tau}  
    \end{aligned}
\end{equation}
In Eq. \ref{eq:first_contribution_of_electronic_possibilities_to_excitons}, the first term represents the excitonic occupation and occurs only in externally induced non-equilibrium, for instance in an optically pumped semiconductor discussed in Sec. \ref{sec:optically_pumped_excitonic_spectrum}. The second term results from pair - anti-pair vacuum fluctuations and involves excitonic properties such as the exciton wavefunction $\varphi^{\nu}_{\boldsymbol{k_1}}$ and the exciton energy $\hbar \Omega^{\nu}_{\Delta\boldsymbol{q}}$. The occuring correlation function Eq. \ref{eq:correlation_function_with_use_of_spatial_integrals} without optical pumping is denoted by $G^{vccv}_{0}(\Delta\boldsymbol{q},t\!+\! \tau,t)$, with the subscript $0$ denoting the intrinsic, unpumped situation:
\begin{widetext}
    \begin{equation}
\label{eq:G_0_vccv}
\begin{aligned}
    G^{vccv}_{0}(\Delta\boldsymbol{q},t\!+\! \tau,t)= \left|\boldsymbol{A}^{\! \perp}_{\! X} \omega_{X}\! \frac{ r_{0}}{ R }\right|^2 \! \sum_{\substack{\boldsymbol{k_1}, \boldsymbol{k_2}\\ \nu }} \! \varphi^{\nu}_{\boldsymbol{k_1}+ \beta \Delta\boldsymbol{q}} \varphi^{*\,\nu}_{ \boldsymbol{k_2}+\beta \Delta\boldsymbol{q}}   e^{i(\omega_{X}-\Omega^{\nu}_{\Delta\boldsymbol{q}}) \tau}  e^{\!-\frac{\gamma^{\nu}_{\Delta\boldsymbol{q}}}{2}\tau} 
\braket{v,\boldsymbol{k_1}|c, \boldsymbol{k_1}\!+\!\Delta \boldsymbol{q}}  \braket{c,\boldsymbol{k_2}\!+\! \Delta\boldsymbol{q}|v, \boldsymbol{k_2}}
    \end{aligned}
\end{equation}
\end{widetext}
Eq. \ref{eq:G_0_vccv} is valid for both, Wannier- and Frenkel excitons. Whereas Wannier excitons are delocalized  over many elementary cells in real space, Frenkel excitons are tightly localized. Frenkel excitons have already been observed in momentum and energy resolved inelastic x-ray scattering from an unpumped crystal \cite{abbamonte_dynamical_2008}. The related features of these excitons are contained in Eq. \ref{eq:G_0_vccv}, which can be reformulated, e.g. in terms of a linear combination of atomic orbitals (LCAO) expansion of the Bloch integrals.
In the opposite limit for Wannier excitons which are the focus of this work, i.e. for $\Delta \boldsymbol{q}$ restricted to the scale of the inverse exciton Bohr radius $a_B^{-1}$, we expand the overlap matrix elements in $\boldsymbol{k}\cdot \boldsymbol{p}$ perturbation theory\cite{haug_quantum_2009} :

\begin{equation}
    \ket{\lambda,\boldsymbol{k}\!+\!\Delta\boldsymbol{q}} \approx   \ket{\lambda,\boldsymbol{k}} - \frac{\hbar}{m}\Delta\boldsymbol{q}\cdot \! \sum_{\lambda' \neq \lambda} \frac{\braket{\lambda', \boldsymbol{k}| \boldsymbol{p} |\lambda, \boldsymbol{k}}}{\epsilon^{\lambda'}_{\boldsymbol{k}} - \epsilon^{\lambda}_{\boldsymbol{k}}} \ket{\lambda', \boldsymbol{k}}
\end{equation}
and make use of the relation between the momentum matrix element $p^{\lambda',\lambda}_{\boldsymbol{k},\boldsymbol{k}}=  \braket{\lambda', \boldsymbol{k}| \boldsymbol{p} |\lambda, \boldsymbol{k}}  $ and the position matrix element $\boldsymbol{r}^{\lambda', \lambda}_{\boldsymbol{k},\boldsymbol{k}}=\braket{\lambda', \boldsymbol{k}| \boldsymbol{r} |\lambda, \boldsymbol{k}} $:
\begin{equation}
\begin{aligned}
       p^{\lambda',\lambda}_{\boldsymbol{k},\boldsymbol{k}}    =\frac{im}{\hbar } (\epsilon^{\lambda'}_{\boldsymbol{k}}-\epsilon^{\lambda}_{\boldsymbol{k}})\boldsymbol{r}^{\lambda', \lambda}_{\boldsymbol{k},\boldsymbol{k}}.
\end{aligned}
\end{equation}
Therefore we approximate for Wannier excitons:
\begin{equation}
\label{eq:kp-theory-approximation}
    \braket{v,\boldsymbol{k_1}|c, \boldsymbol{k_1}\! + \! \Delta \boldsymbol{q}}  \approx -i \Delta\boldsymbol{q} \cdot \boldsymbol{r}^{v,c}_{\boldsymbol{k_1},\boldsymbol{k_1}}
\end{equation}
and the correlation function $G^{vccv}_{0}$ can be written as:
\begin{equation}
\label{eq:eq:G_0_vccv_with_kp}
\begin{aligned}
    G^{vccv}_{0}(\Delta\boldsymbol{q},t\!+\! \tau,t)&= \! \biggl|\boldsymbol{A}^{\! \perp}_{\! X} \omega_{X}\! \frac{ r_{0}}{ R }\biggr|^2 \hspace{-0.1cm} \sum_{\nu} \biggl|\Delta\boldsymbol{q} \cdot \boldsymbol{r}_{v,c}\! \sum_{\boldsymbol{k}_{1}} \!\varphi^{\nu}_{\boldsymbol{k}_{1}}  \biggr|^2 
    \\
    &\times e^{i(\omega_{X}-\Omega^{\nu}_{\Delta\boldsymbol{q}}) \tau} e^{\!-\frac{\gamma^{\nu}_{\Delta\boldsymbol{q}}}{2}\tau}  
    \end{aligned}
\end{equation}
where $\boldsymbol{r}_{v,c} = \braket{v,\boldsymbol{0}|\boldsymbol{r}|c,\boldsymbol{0}}$. \\
Similarly, we calculate the contribution from the third line in Eq. \ref{eq:possibilities_electronic_operators}:
\begin{equation}
\label{eq:G_0_vvvv}
\begin{aligned}
    G^{vvvv}_{0}(\Delta\boldsymbol{q},t\!+\! \tau,t)&= \left|\boldsymbol{A}^{\! \perp}_{\! X} \omega_{X}\! \frac{ r_{0}}{ R }\right|^2  \biggl|\sum_{\boldsymbol{k_1} }\! \braket{v,\boldsymbol{k_1}|v, \boldsymbol{k_1}} \!\biggr|^2  
    \\
    &\times e^{i\omega_{X} \tau} e^{\!-\frac{\gamma^{\nu}_{\Delta\boldsymbol{q}}}{2}\tau}  \delta_{\Delta\boldsymbol{q},\boldsymbol{0}}
    \end{aligned}
\end{equation}
The sum appearing in Eq. \ref{eq:G_0_vvvv} is equivalent to the number of unit cells $N$ in the crystal\cite{czycholl_theoretische_2016}:
\begin{equation}
\begin{aligned}
    \sum_{\boldsymbol{k_1} } \braket{v,\boldsymbol{k_1}|v, \boldsymbol{k_1}}
    = \sum_{\boldsymbol{k_1} } = A    \int_{BZ}\frac{ d^2 k}{4\pi^2} = A\frac{A_{BZ}}{4\pi^2}       = N
\end{aligned}
\end{equation}
Here, $A_{BZ}$ is the 2D Brillouin zone volume and we used that $ \frac{A_{BZ}}{4\pi^2} = \frac{1}{A_{UC}}$. 
For a better comparison of both contributions Eq. \ref{eq:G_0_vccv}, \ref{eq:G_0_vvvv} to the intrinsic spectrum we rewrite the sum over the excitonic wavefunction in Eq. \ref{eq:eq:G_0_vccv_with_kp}:
\begin{equation}
\label{eq:rearranging_sum_over_excitonic_wavefunction}
\begin{aligned}
        \sum_{\boldsymbol{k}_{1}} \varphi^{\nu}_{\boldsymbol{k}_{1}} &= A \varphi^{\nu}(\boldsymbol{r=0}) = N A_{UC} \varphi^{\nu}(\boldsymbol{r=0}).
        \end{aligned}
\end{equation}
As explained in Appendix \ref{app:fourier_transform_and_Excitons}, we choose to use unitless wavefunctions in reciprocal space $\varphi^{\nu}_{\boldsymbol{k}}$. Therefore, the Fourier transformed real space wavefunction $\varphi^{\nu}(\boldsymbol{r})$ has the unit $\si{\meter^{-2}}$ and reads $\varphi^{\nu}(\boldsymbol{r}) = \frac{\tilde{\varphi}^{\nu}(\boldsymbol{r})}{a_{B}^2}$ with the Bohr radius of the exciton $a_{B}$ and the unitless wavefunction $\tilde{\varphi}^{\nu}(\boldsymbol{r})$ localized over the Bohr radius of the Wannier exciton, cp. Eq. \ref{eq:app:fourier_transform_exciton_wavefunction_bohr_radius}. In this notation Eq. \ref{eq:rearranging_sum_over_excitonic_wavefunction} reads:
\begin{equation}
\begin{aligned}
        \sum_{\boldsymbol{k}_{1}} \varphi^{\nu}_{\boldsymbol{k}_{1}} = N \frac{A_{UC}}{a_{B}^2} \tilde{\varphi}^{\nu}(\boldsymbol{r=0}) 
        \end{aligned}
\end{equation}

The squared Bohr radius $a_{B}^2$ can be interpreted as a measure of the exciton extension and therefore the term $\frac{A_{UC}}{a_{B}^2}$ is proportional to the spatial fraction of the total exciton distributed per unit cell. In the case of spatially delocalized Wannier excitons this value is much smaller than one. In the following, we will use the semiconductor WS$_2$ as an example where $\frac{A_{UC}}{a_{B}^2}\approx 0.025$. \\ 
Collecting all contributions, we find for the intrinsic x-ray, i.e. unpumped spectrum $ S_{0}(\Delta\boldsymbol{q},\omega)$:
\begin{widetext}
    \begin{equation}
    \label{eq:unpumped_spectrum}
    \begin{aligned}
 S_{0}(\Delta\boldsymbol{q},\omega) =& \left|\boldsymbol{A}^{\! \perp}_{\! X} \omega_{X}\! \frac{ r_{0}}{ R }N \right|^2  2\pi \delta(\omega-\omega_{X})\delta_{\Delta\boldsymbol{q},\boldsymbol{0}}
 \\
 +& \left|\boldsymbol{A}^{\! \perp}_{\! X} \omega_{X}\! \frac{ r_{0}}{ R }N\right|^2 \sum_{\nu} \biggl|\Delta\boldsymbol{q} \cdot \boldsymbol{r}_{v,c}  \tilde{\varphi}^{\nu}(\boldsymbol{r=0}) \frac{A_{UC}}{a_{B}^2} \biggr|^2   \frac{\gamma^{\nu}_{\Delta\boldsymbol{q}}}{\bigl(\frac{\gamma^{\nu}_{\Delta\boldsymbol{q}}}{2}\bigr)^2 + (\omega-\omega_{X}+\Omega^{\nu}_{\Delta\boldsymbol{q}})^2} . 
    \\
    \end{aligned}
\end{equation}
\end{widetext}
Due to the radiative and exciton-phonon induced dephasing, linewidth-broadened Lorentzians are obtained for the lineshape functions. The first contribution constitutes the elastic scattering of the valence band electrons resulting from $G^{vvvv}_{0}$.
The second contribution, resulting from $G^{vccv}_{0}$, contains the sum over all excitonic resonances $\Omega^{\nu}_{\Delta\boldsymbol{q}}$. 
The resonances occuring in the spectrum, Eq. \ref{eq:unpumped_spectrum}, allows to study Wannier exciton dispersions $\Omega^{\nu}_{\Delta\boldsymbol{q}}$ in x-ray scattering.\\
\begin{figure*}[ht!]
    \centering
    \begin{subfigure}[b]{0.49\textwidth}
            \centering
    \includegraphics[width=\linewidth]{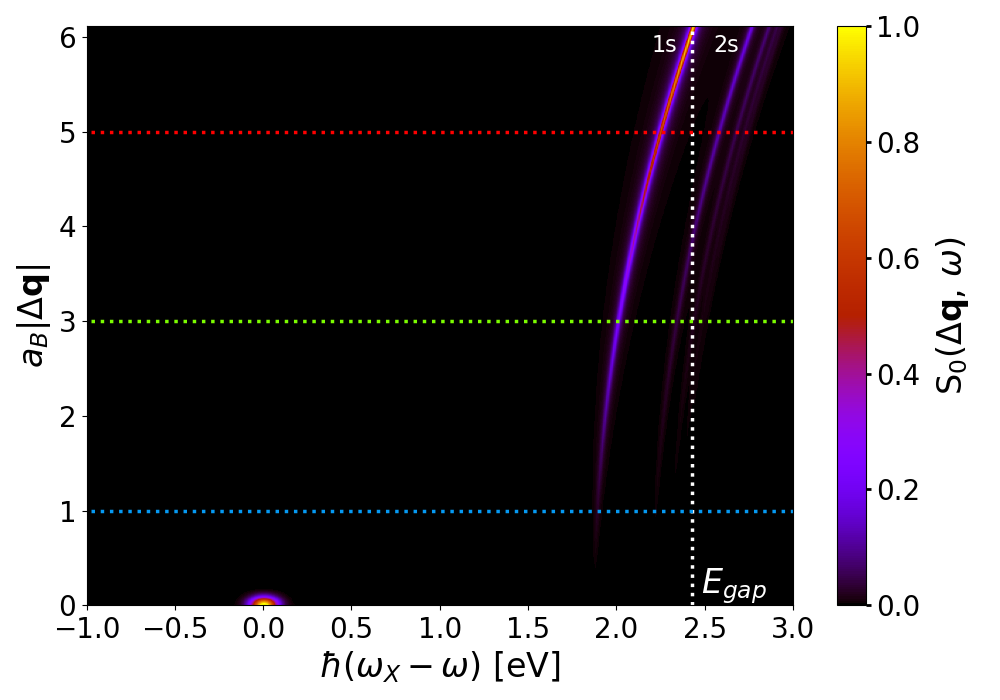}
            \caption{}
            \label{fig:spectrum_unpumped_for_WS2_a}
    \end{subfigure}%
    \begin{subfigure}[b]{0.49\textwidth}
            \centering
                \includegraphics[width=\linewidth]{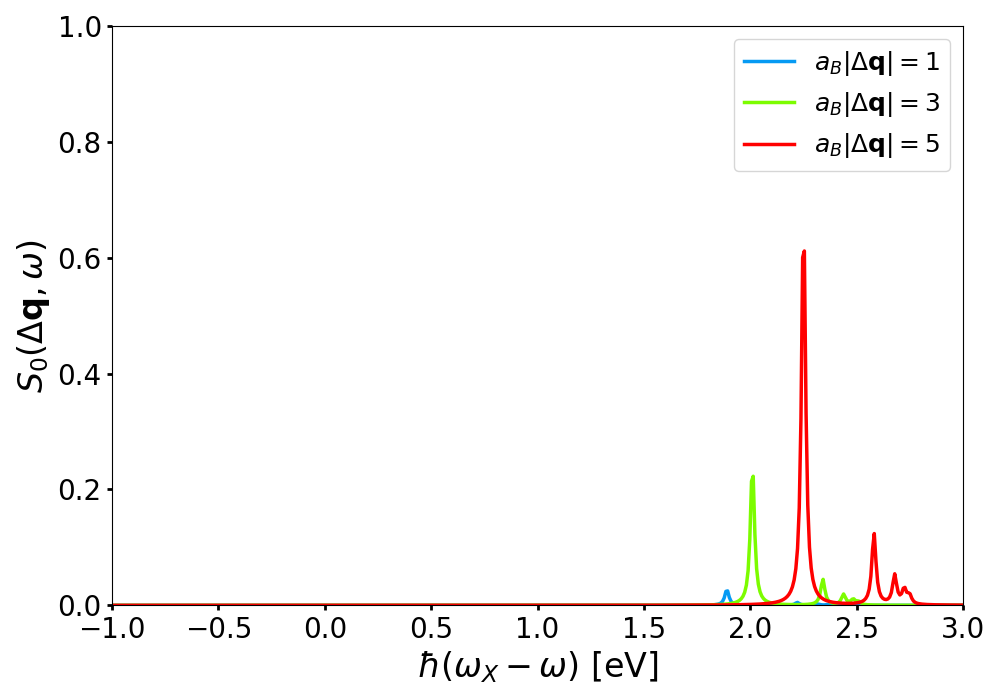}
            \caption{}
            \label{fig:spectrum_unpumped_for_WS2_b}
    \end{subfigure}
    \caption{Intrinsic x-ray scattering spectra for the material WS$_2$ (without optical pumping). Due to reasons of visibility the elastic scattering contribution was normalized with respect to the highest value of the excitonic contribution and we used also the same linewidth as for the excitons. The spectrum is normalized with respect to the maximum value.  (a) Energy loss and momentum transfer dependent spectrum. The vertical white dashed line denotes the band gap energy while the horizontally dashed lines are cuts of the spectrum which are shown in (b). To improve comparability, the colors used in (b) match the colors of the cuts (horizontal dashed lines).}
      \label{fig:spectrum_unpumped_for_WS2}
\end{figure*}
To illustrate Eq. \ref{eq:unpumped_spectrum}, Fig. \ref{fig:spectrum_unpumped_for_WS2_a} shows the spectrum of an intrinsic  WS$_2$ monolayer as a function of momentum transfer scaled with respect to the exciton Bohr radius $a_B |\Delta\boldsymbol{q}|$ and energy transfer  $\hbar(\omega_{X}\! -\! \omega)$. The material parameters used for the evaluation of Eq. \ref{eq:unpumped_spectrum} are provided in table \ref{tab:values_of_effective_masses_and_binding_energies} in App. \ref{app:material_parameters}. The spectrum incorporates a linewidth of $\hbar\gamma =\SI{23.2}{meV}$ which is in alignment with the results obtained from Selig et al. \cite{selig_excitonic_2016} by microscopically calculating exciton-phonon and radiative linewidth contributions at $\SI{300}{\kelvin}$. 
At vanishing momentum transfer $a_B|\Delta\boldsymbol{q}| = 0$ and vanishing energy transfer $\hbar(\omega_{X}\! -\! \omega)=0$, the well known elastic Thomson scattering contribution of the valence band is observed. This peak has been normalized to the highest value of the excitonic contribution, which scales as $A_{UC}/a_{B}^2$, cp. Eq. \ref{eq:unpumped_spectrum}.
Since x-ray scattering for the intrinsic semiconductor, Eq. \ref{eq:unpumped_spectrum}, exhibits only Wannier exciton states with $\varphi^{\nu}(\boldsymbol{r}=\boldsymbol{0}) $, only s-states appear in Fig. \ref{fig:spectrum_unpumped_for_WS2}. We note that these features are identical to the excitonic states seen in optical spectra\cite{kira_semiconductor_2011}. The most dominant contribution can be assigned to the 1s exciton, the next contribution (further right in the spectrum) can be assigned to the 2s exciton and so on.
Note, that only finite scattering angles $\Delta\boldsymbol{q}\neq \boldsymbol{0}$ contribute to the excitonic signal increasing with momentum transfer, cp. Eq. \ref{eq:unpumped_spectrum}. The spectral peaks in Fig. \ref{fig:spectrum_unpumped_for_WS2_a} follow the energy dispersion of the exciton. A similar behaviour of an increased excitonic signal for a larger momentum transfer was as well observed in \cite{hamalainen_momentum_2002,abbamonte_dynamical_2008} for the material of LiF. However, as noted above the LiF excitons are Frenkel excitons and their x-ray scattering spectra exhibit different information in comparison to Wannier excitons.
In Fig \ref{fig:spectrum_unpumped_for_WS2_b} we show spectral traces from \ref{fig:spectrum_unpumped_for_WS2_a} for different values of $a_{B}|\Delta\boldsymbol{q}|$. The colors of the cuts in Fig. \ref{fig:spectrum_unpumped_for_WS2_a} have been chosen to match the colors in Fig. \ref{fig:spectrum_unpumped_for_WS2_b}. Fig. \ref{fig:spectrum_unpumped_for_WS2_b} shows more clearly that with increasing momentum transfer also the signal strength increases. While for $a_{B}|\Delta\boldsymbol{q}|=1$ the excitonic signal is almost imperceptible, all contributions rise in intensity when increasing the momentum transfer. The ratio of the contributions from different excitonic states remains constant as $\Delta\boldsymbol{q}$ is varied.\\

\subsection{Optically pumped excitonic spectrum}
\label{sec:optically_pumped_excitonic_spectrum}
To determine the x-ray scattering spectrum under resonant optical pumping of excitons, all contributions of the type $ \braket{ P^{\dagger\, \nu'}_{\! \Delta\boldsymbol{q}}(t)  P^{ \nu}_{\! \Delta\boldsymbol{q}}(t\!+\! \tau) } $, cp. Eq. \ref{eq:pair-operators_to_exciton_operators}, occurring in Eq. \ref{eq:possibilities_electronic_operators} must be considered.
As an example, we discuss the remaining first term of $ G^{vccv}$, occurring on the right hand side in Eq. \ref{eq:first_contribution_of_electronic_possibilities_to_excitons}:
\begin{widetext}
\begin{equation}
\label{eq:correlation_function_G_vccv_vanish}
\begin{aligned}
    G^{vccv}_{\text{pump}}(\Delta\boldsymbol{q},t\!+\!\tau,t)= \left|\boldsymbol{A}^{\! \perp}_{\! X} \omega_{X}\! \frac{ r_{0}}{ R }\right|^2 &\sum_{\substack{\boldsymbol{k_1}, \boldsymbol{k_2} \\\nu,\nu'}}\varphi^{*\,\nu'}_{ \boldsymbol{k_2}+\beta \Delta\boldsymbol{q}}\varphi^{\nu}_{\boldsymbol{k_1}+ \beta \Delta\boldsymbol{q}} \braket{ P^{\dagger\, \nu'}_{\! \Delta\boldsymbol{q}}(t)  P^{ \nu}_{\! \Delta\boldsymbol{q}}(t\!+\! \tau) }\ e^{i\omega_{X} \tau}
      \\
     & \times \braket{v,\boldsymbol{k_1}|c, \boldsymbol{k_1}\! + \! \Delta \boldsymbol{q}}  \braket{c,\boldsymbol{k_2}+\Delta\boldsymbol{q}|v, \boldsymbol{k_2}}.
    \end{aligned}
\end{equation}
\end{widetext}
Vanishing momentum transfer by the optical field implies $ \braket{ P^{\dagger\, \nu'}_{\! \Delta\boldsymbol{q}}(t)  P^{ \nu}_{\! \Delta\boldsymbol{q}}(t\!+\! \tau) } \approx  \braket{ P^{\dagger\, \nu'}_{\boldsymbol{0}}(t)  P^{ \nu}_{\boldsymbol{0}}(t\!+\! \tau) } \delta_{\Delta \boldsymbol{q},\boldsymbol{0}}$. 
Therefore, the correlation function, Eq. \ref{eq:correlation_function_G_vccv_vanish}, vanishes for Wannier excitons due to the orthogonality of the Bloch factors $\braket{v,\boldsymbol{k_1}|c, \boldsymbol{k_1}} = 0$:
\begin{equation}
\begin{aligned}
    G^{vccv}_{\text{pump}}(\Delta\boldsymbol{q},t\!+\! \tau,t)= 0
    \end{aligned}
\end{equation}
In fact, calculating all remaining contributions of Eq. \ref{eq:possibilities_electronic_operators} to lowest order in the optical pump and in $\Delta \boldsymbol{q}$ occurring on the length scale of the inverse excitonic Bohr radius, one finds that only the following correlations of electron and hole densities contribute:
\begin{equation}
\begin{aligned}
         &G_{pump}(\Delta\boldsymbol{q},t\!+\! \tau,t)
         \\
         =&G^{vvvv}_{pump}(\Delta\boldsymbol{q},t\!+\! \tau,t)
+G^{cccc}_{pump}(\Delta\boldsymbol{q},t\!+\! \tau,t)
         \\
         +&G^{ccvv}_{pump}(\Delta\boldsymbol{q},t\!+\! \tau,t)
+G^{vvcc}_{pump}(\Delta\boldsymbol{q},t\!+\! \tau,t).
\\
         \end{aligned}
\end{equation}
For instance, the two-time correlation contribution from $G^{cccc}(\Delta\boldsymbol{q},t\!+\! \tau,t)$ reads:
\begin{widetext}
    \begin{equation}
    \begin{aligned}
        &\braket{c^{\dagger}_{\boldsymbol{k_1}}(t\!+\! \tau) c_{\boldsymbol{k_1}+\Delta\boldsymbol{q}}(t\!+\! \tau)  c^{\dagger}_{\boldsymbol{k_2}+\Delta\boldsymbol{q}}(t) c_{\boldsymbol{k_2}}(t)}
        \\
    =&\sum_{\substack{\boldsymbol{k'},\boldsymbol{k''}\\ \nu,\nu' \\ \mu,\mu'}}\varphi^{*\,\nu}_{ \beta \boldsymbol{k_1}+\alpha\boldsymbol{k'}} \varphi^{\nu'}_{ \beta (\boldsymbol{k_1} + \Delta\boldsymbol{q)}+\alpha\boldsymbol{k'}} \varphi^{*\,\mu}_{ \beta (\boldsymbol{k_2} + \Delta\boldsymbol{q})+\alpha\boldsymbol{k''}} \varphi^{\mu'}_{ \beta \boldsymbol{k_2}+\alpha\boldsymbol{k''}}
    \\
    \times& \braket{P^{\dagger \, \nu}_{\boldsymbol{k_1}-\boldsymbol{k'}} (t\!+\! \tau)  P^{\nu'}_{\boldsymbol{k_1}+\Delta\boldsymbol{q}-\boldsymbol{k'}} (t\!+\! \tau)P^{\dagger \, \mu}_{\boldsymbol{k_2} + \Delta\boldsymbol{q}-\boldsymbol{k''}} (t)  P^{\mu'}_{\boldsymbol{k_2}-\boldsymbol{k''}} (t)}
    \end{aligned}
\end{equation}
\end{widetext}
Normal ordering of the excitonic operators by using the commutator in Eq. \ref{eq:generalized_commutator} yields:
    \begin{equation}
\begin{aligned}
    &\braket{P^{\dagger \, \nu}_{\boldsymbol{k_1}-\boldsymbol{k'}} (t\!+\! \tau)  P^{\nu'}_{\boldsymbol{k_1}+\Delta\boldsymbol{q}-\boldsymbol{k'}} (t\!+\! \tau)P^{\dagger \, \mu}_{\boldsymbol{k_2} + \Delta\boldsymbol{q}-\boldsymbol{k''}} (t)  P^{\mu'}_{\boldsymbol{k_2}-\boldsymbol{k''}} (t)} 
    \\
    =& \braket{P^{\dagger \, \nu}_{\boldsymbol{k_1}-\boldsymbol{k'}} (t\!+\! \tau)    P^{\mu'}_{\boldsymbol{k_2}-\boldsymbol{k''}} (t)} \delta^{\nu', \mu}_{\boldsymbol{k_1},\boldsymbol{k_2}} e^{\!-i\Omega^{\nu'}_{\boldsymbol{k_1}+\Delta\boldsymbol{q}-\boldsymbol{k'}} \tau}e^{\!-\frac{\gamma^{\nu'}_{\boldsymbol{k_1}+\Delta\boldsymbol{q}-\boldsymbol{k'}}}{2}\tau} 
    \\
    +&\braket{P^{\dagger \, \nu}_{\boldsymbol{k_1}-\boldsymbol{k'}} (t\!+\! \tau)  P^{\dagger \, \mu}_{\boldsymbol{k_2} + \Delta\boldsymbol{q}-\boldsymbol{k''}} (t) P^{\nu'}_{\boldsymbol{k_1}+\Delta\boldsymbol{q}-\boldsymbol{k'}} (t\!+\! \tau) P^{\mu'}_{\boldsymbol{k_2}-\boldsymbol{k''}} (t)}
\end{aligned}
\end{equation}
In the low-density limit, i.e. weak optical pumping, only the first contribution has to be taken into account: 
\begin{widetext}
        \begin{equation}
\label{eq:G_cccc}
\begin{aligned}
    G^{cccc}_{pump}(\boldsymbol{r};\Delta\boldsymbol{q},t\!+\! \tau,t)= \left|\boldsymbol{A}^{\! \perp}_{\! X} \omega_{X}\! \frac{ r_{0}}{ R }\right|^2 &\sum_{\substack{\boldsymbol{k_1}, \boldsymbol{k_2} \\ \nu,\nu',  \mu'}} \varphi^{*\,\nu}_{\boldsymbol{k_1}}  \varphi^{\nu'}_{  \boldsymbol{k_1}+ \beta \Delta\boldsymbol{q}} \varphi^{*\,\nu'}_{  \boldsymbol{k_2} +\beta \Delta\boldsymbol{q}}  \varphi^{\mu'}_{ \boldsymbol{k_2}} \braket{ P^{\dagger \, \nu}_{\boldsymbol{0}} (t\!+\! \tau) P^{\mu'}_{\boldsymbol{0}} (t)} e^{i(\omega_{X}-\Omega^{\nu'}_{\Delta\boldsymbol{q}} ) \tau}
 e^{\!-\frac{\gamma^{\nu'}_{\Delta\boldsymbol{q}}}{2}\tau}  
 \\
    &  \times \braket{c,\boldsymbol{k_1}|c, \boldsymbol{k_1}\! + \! \Delta \boldsymbol{q}}  \braket{c,\boldsymbol{k_2}+\Delta\boldsymbol{q}|c, \boldsymbol{k_2}}
    \end{aligned}
\end{equation}
\end{widetext}
1In the steady-state limit of $\braket{ P^{\dagger \,\nu}_{\boldsymbol{Q}}(t\!+\! \tau) P^{\mu}_{\boldsymbol{Q'}}(t)}$ we find the relation:
\begin{widetext}

        \begin{equation}
 \label{eq:correlation_function_excitons_long_time_limit}
    \begin{aligned}
     \lim_{t\to\infty} \!  \braket{ P^{\dagger \,\nu}_{\boldsymbol{Q}}(t\!+\! \tau) P^{\mu}_{\boldsymbol{Q'}}(t)}   
     =&  \frac{ \delta_{\boldsymbol{Q},\boldsymbol{0}} \delta_{\boldsymbol{Q'},\boldsymbol{0}} (\boldsymbol{\Omega}^{\mu}_{c,v}\! \cdot\! \boldsymbol{E}_{o}) (\boldsymbol{\Omega}^{\nu}_{v,c}\! \cdot\! \boldsymbol{E}_{o})  }{{\bigl(\frac{\gamma^{\mu}_{\boldsymbol{Q'}}}{2}\! + i(\Omega^{\mu}_{\boldsymbol{Q'}}\!-\!\omega_{o}) \bigr)\bigl(\frac{\gamma^{\nu}_{\boldsymbol{Q}}}{2}\! -\! i(\Omega^{\nu}_{\boldsymbol{Q}}\!-\omega_{o})\bigr) } }e^{i\omega_{o}  \tau}.
    \end{aligned}
\end{equation}

\end{widetext}
which can be inserted into Eq. \ref{eq:G_cccc}.
All other contributions occurring in Eq. \ref{eq:possibilities_electronic_operators} are calculated in a similar manner and we obtain for the optically pumped contribution $S_{\text{pump}}(\Delta\boldsymbol{q},\omega)$ of the x-ray scattering spectrum:
\begin{widetext}
    \begin{equation}
\label{eq:pumped_spectrum_final}
\begin{aligned}
S_{\text{pump}}(\Delta\boldsymbol{q},\omega)
\!=&\left|\boldsymbol{A}^{\! \perp}_{\! X} \omega_{X}\! \frac{ r_{0}}{ R }\right|^2  \!\sum_{\nu'} \biggl| \sum_{\substack{\nu}}\! \frac{ (\boldsymbol{\Omega}^{\nu}_{v,c} \!\cdot\! \boldsymbol{E}_{o})  }{{\bigl(\frac{\gamma^{\nu}_{\boldsymbol{0}}}{2}\! - i(\Omega^{\nu}_{\boldsymbol{0}}-\omega_{o})\bigr) } }\! \sum_{\boldsymbol{k_1}}\bigl( \varphi^{*\,\nu}_{ \boldsymbol{k_1}+\alpha\Delta\boldsymbol{q}}\varphi^{\nu'}_{\boldsymbol{k_1}} -\varphi^{*\,\nu}_{  \boldsymbol{k_1}} \varphi^{\nu'}_{\boldsymbol{k_1}+ \beta \Delta\boldsymbol{q}}\bigr) \biggr|^2 \!\frac{\gamma^{\nu'}_{\Delta\boldsymbol{q}}}{\bigl(\frac{\gamma^{\nu'}_{\Delta\boldsymbol{q}}}{2}\bigr)^2 \! + \! (\omega_{X}\! -\! \Omega^{\nu'}_{\Delta\boldsymbol{q}}\! +\! \omega_{o}\! -\! \omega)^2}.
    \end{aligned}
\end{equation}
\end{widetext}
In experimental pump-probe geometry, the pump induced part of the spectrum Eq. \ref{eq:pumped_spectrum_final}, can be obtained by subtracting the spectrum without pumping from the full, optically pumped scattering spectrum $S(\Delta\boldsymbol{q},\omega)$, i.e. in a difference scattering spectrum (DSS):
\begin{equation}
    S_{\text{pump}}(\Delta\boldsymbol{q},\omega) = S(\Delta\boldsymbol{q},\omega) - S_{0}(\Delta\boldsymbol{q},\omega)
\end{equation}
A key feature of $S_{\text{pump}}(\Delta\boldsymbol{q},\omega)$, Eq. \ref{eq:pumped_spectrum_final} is given by the convolution of exciton wavefunctions: $\sum_{\boldsymbol{k_1}}\bigl( \varphi^{*\,\nu}_{ \boldsymbol{k_1}+\alpha\Delta\boldsymbol{q}}\varphi^{\nu'}_{\boldsymbol{k_1}} -\varphi^{*\,\nu}_{  \boldsymbol{k_1}} \varphi^{\nu'}_{\boldsymbol{k_1}+ \beta \Delta\boldsymbol{q}}\bigr)$. 
This contribution is only non-vanishing for finite wave-vector transfer, i.e. $a_B|\Delta\boldsymbol{q}|\neq 0$. For the case $\nu=\nu'$ one additionally needs a spatial imbalance for the internal electron and hole charge distribution, i.e. $\alpha\neq \beta$, defined after Eq. \ref{eq:time_evolution_of_electron_hole_pair_operator}. Additionally, the excitonic wavefunction contribution contains all interferences of optically excited excitons $\nu$ (s-states) with all excitonic resonances $\nu'$, including optically forbidden resonances (non-s-like states).
In particular, when neglecting quantum interferences between different excitonic states, i.e. for $\nu\!=\!\nu'$ the convolution of the exciton wavefunctions is the Fourier transform of the total charge distribution of the $\nu^{\text{th}}$-exciton state: $\rho^{\nu}_{tot}(\boldsymbol{r})\! =\!\rho^{\nu}_{h}(\boldsymbol{r}) \!- \!\rho^{\nu}_{el}(\boldsymbol{r})$, given as the difference of the charge distribution of the associated hole $\rho^{\nu}_{ho}(\boldsymbol{r})$ and electron $\rho^{\nu}_{el}(\boldsymbol{r})$. The possibility to determine the internal charge distribution of the exciton from the x-ray scattering spectrum will be discussed explicitely in section \ref{sec:charge_distribution_of_exciton}.\\
Here, we note that by defining the quantity $  \varrho^{\dagger\hspace{0.025cm}\nu'} \!(\Delta\boldsymbol{q},t)$:
\begin{equation}
\label{eq:new_quantity_for_dynamic_sturcture_factor}
\begin{aligned}
    \varrho^{\dagger\hspace{0.025cm}\nu'} \!(\Delta\boldsymbol{q},t) \! &=\!\sum_{\substack{\nu,\boldsymbol{k_1}}} \!P^{\dagger\hspace{0.025cm} \nu}_{\boldsymbol{0}}\!(t)e^{\!-i\Omega^{\nu'}_{\Delta\boldsymbol{q}}t} \bigl(\varphi^{\nu'}_{\boldsymbol{k_1}}\! \varphi^{*\,\nu}_{ \boldsymbol{k_1}+\alpha\Delta\boldsymbol{q}} \!-\!\varphi^{*\,\nu}_{  \boldsymbol{k_1}}\! \varphi^{\nu'}_{\boldsymbol{k_1}+ \beta \Delta\boldsymbol{q}}\bigr), 
    \end{aligned}
\end{equation}
the pumped scattering spectrum can be written as a dynamic structure factor involving also excitonic coherences $\nu\neq \nu'$:
\begin{widetext}
    \begin{equation}
\label{eq:pumped_spectrum_final_alternate_form}
\begin{aligned}
S_{\text{pump}}(\Delta\boldsymbol{q},\omega)
=&\left|\boldsymbol{A}^{\! \perp}_{\! X} \omega_{X}\! \frac{ r_{0}}{ R }\right|^2 \! \frac{1}{2\pi}\!  \int dt e^{i(\omega_{X} - \omega)t} \sum_{\nu'} \braket{\varrho^{\dagger\hspace{0.025cm}\nu'} \!(\Delta\boldsymbol{q},t)\varrho^{\nu'}(\Delta\boldsymbol{q},0)}
    \\
    \end{aligned}
\end{equation}
\end{widetext}
if we apply the limit $\gamma^{\nu'}_{\Delta\boldsymbol{q}}\to 0$ and the identity:
\begin{equation}
\label{eq:limit_lorentzian_to_fourier_space}
\begin{aligned}
        &\lim_{\gamma^{\nu'}_{\Delta\boldsymbol{q}}\!\to 0}\! \frac{\,\gamma^{\nu'}_{\Delta\boldsymbol{q}}}{\bigl(\frac{\gamma^{\nu'}_{\Delta\boldsymbol{q}}}{2}\bigr)^2\!\! +\! (\omega_{X}\! -\! \Omega^{\nu'}_{\Delta\boldsymbol{q}}\! +\!\omega_{o}\! -\! \omega)^2}\! 
    \\
    =& \frac{1}{2\pi }\!\!\int\! dt e^{i(\omega_{X}\! - \Omega^{\nu'}_{\Delta\boldsymbol{q}}\! +\omega_{o}\! - \omega)t}
    \end{aligned}
\end{equation}
A more thorough examination of the pump induced spectrum, Eq. \ref{eq:pumped_spectrum_final}, will be conducted numerically, applying again the parameters provided in table \ref{tab:values_of_effective_masses_and_binding_energies} for atomically thin WS$_2$.
    \begin{figure*}[!ht]
    \centering
    \begin{subfigure}[b]{0.49\textwidth}
            \centering
    \includegraphics[width=\linewidth]{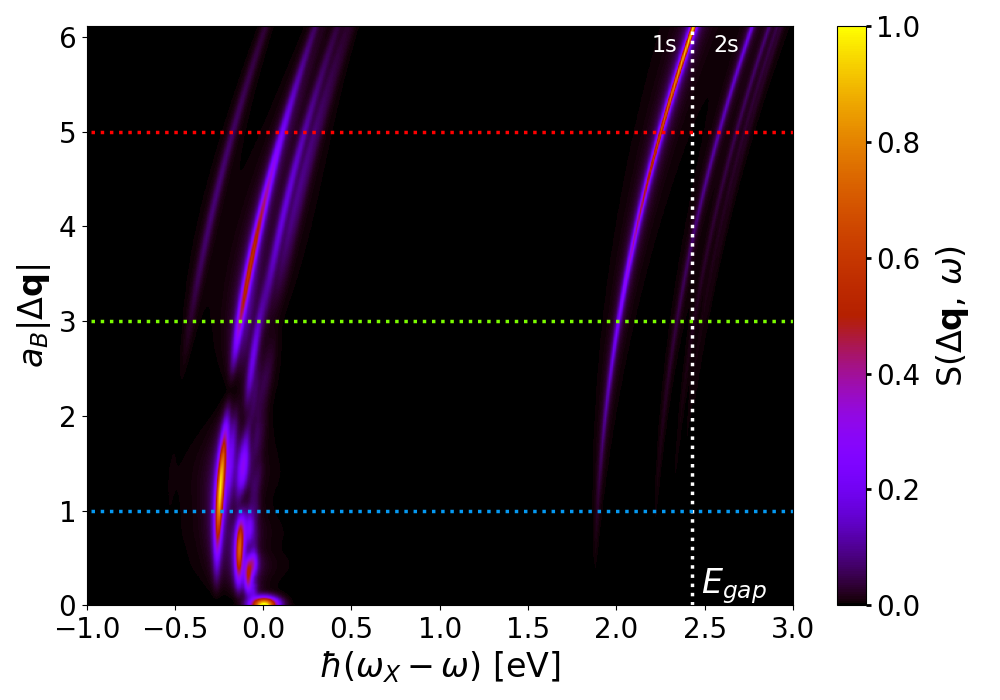}
            \caption{}
    \end{subfigure}%
    \begin{subfigure}[b]{0.49\textwidth}
            \centering
    \includegraphics[width=\linewidth]{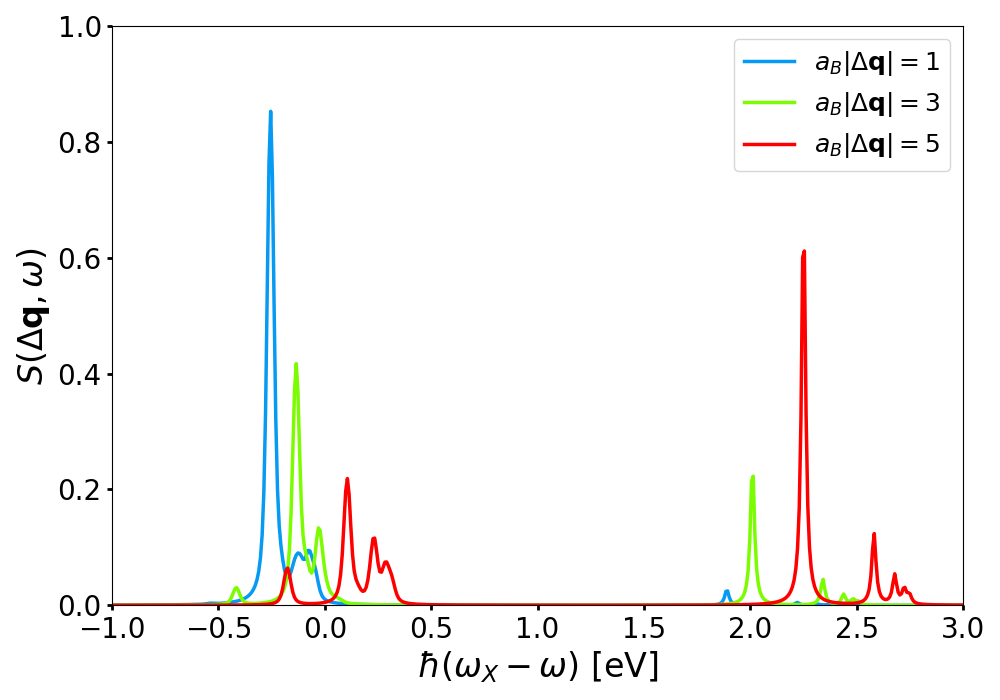}
            \caption{}
    \end{subfigure}
    \caption{Full spectrum $S(\Delta\boldsymbol{q},\omega)$ of WS$_2$ including elastic and inelastic contributions from the pumped semiconductor. The total spectrum is normalized. As in Fig. \ref{fig:spectrum_unpumped_for_WS2} the contribution from the elastic scattering was scaled to the highest value of the excitonic contribution. (a) Energy loss and momentum transfer dependent spectrum. The vertical white dashed line shows the energy of the band gapwhile the horizontally dashed lines are cuts of the spectrum shown in (b). For a better visibility the colors of the dashed lines are matching with the colors used in (b).}
  \label{fig:spectrum_pump_for_WS2}
\end{figure*}
Fig. \ref{fig:spectrum_pump_for_WS2} shows the spectrum $S(\Delta\boldsymbol{q},\omega)$ as a function of the energy transfer $\hbar(\omega_{X}\!\ -\! \omega)$ and the momentum transfer scaled by the exciton Bohr radius $a_B |\Delta\boldsymbol{q}|$ for the optical pump tuned to the band edge of the semiconductor, $\hbar \omega_{o} = E_{\text{gap}}$. In addition to the contributions of the intrinsic inelastic scattering, also obtained for the unpumped system in Fig. \ref{fig:spectrum_unpumped_for_WS2}, we obtain new features close to $\hbar(\omega_{X}\! -\! \omega)\approx 0$ indicating a new quasi-elastic scattering contribution. These features can be traced back to the convolution and interference of the excitonic wavefunctions in Eq. \ref{eq:pumped_spectrum_final} and therefore to the internal excitonic charge distribution Eq. \ref{eq:new_quantity_for_dynamic_sturcture_factor}. 
Due to the dependence of the electron-hole distribution on the effective mass ratios $\alpha$ and $\beta$ of electron and hole, this contribution is therefore strongly dependent on the Bohr radii and the reduced exciton mass.
For new quasi-elastic contributions, an energy shift is observed when the momentum transfer $\Delta \boldsymbol{q}$ is increased. This phenomenon can be attributed, similarly to the result obtained from the intrinsic unpumped spectrum, to the dispersion of the exciton $\Omega^{\nu'}_{\Delta\boldsymbol{q}}$, cp. Eq. \ref{eq:pumped_spectrum_final}. For this reason, we use the term quasi-elastic.
Fig. \ref{fig:spectrum_pump_for_WS2}b shows cuts through Fig. \ref{fig:spectrum_pump_for_WS2}a for the same momentum transfer $\Delta\boldsymbol{q}$ in addition to those obtained for the intrinsic unpumped semiconductor Fig. \ref{fig:spectrum_unpumped_for_WS2}. The optically induced quasi-elastic contributions lose intensity with increasing momentum transfer, exhibiting an opposite behaviour to the intrinsic non-elastic spectral contributions occuring at higher energies. In contrast to the intrinsic spectrum, the ratio of the peaks is not fixed when varying the momentum transfer $\Delta\boldsymbol{q}$.
\subsection{Charge distribution of the exciton}
\label{sec:charge_distribution_of_exciton}
To show the connection between the internal charge distribution of the excitons and the spectrum, cp. Eq. \ref{eq:pumped_spectrum_final} and Fig. \ref{fig:spectrum_pump_for_WS2}, we evaluate the Fourier transform of the convolution of the excitonic wavefunctions:
\begin{equation}
    \sum_{\boldsymbol{k_1}}\bigl( \varphi^{*\,\nu}_{ \boldsymbol{k_1}+\alpha\Delta\boldsymbol{q}}\varphi^{\nu'}_{\boldsymbol{k_1}}\! -\varphi^{*\,\nu}_{  \boldsymbol{k_1}} \varphi^{\nu'}_{\boldsymbol{k_1}+ \beta \Delta\boldsymbol{q}}\bigr)
\end{equation}
In the first term we shift $\boldsymbol{k_1} \to \boldsymbol{k_1}-\alpha\Delta\boldsymbol{q}$, use $a_{B} = a_{0} \epsilon \frac{m_{0}}{\mu}$ and :
\begin{equation}
    \begin{aligned}
        a_{B} \alpha &= a_{0} \epsilon \frac{m_{0}}{\mu} \frac{m_{e}}{m_e + m_h} = a_{0} \epsilon \frac{m_{0}}{m_h} \equiv a_h
        \\
        a_{B} \beta &= a_{0} \epsilon \frac{m_{0}}{\mu} \frac{m_{h}}{m_e + m_h} = a_{0} \epsilon \frac{m_{0}}{m_e} \equiv a_{el}
    \end{aligned}
\end{equation}
\\
The wavefunctions vary intrinsically on the length scale of the excitonic Bohr radius and we denote $\varphi^{\nu}(\boldsymbol{k_1}) \to \varphi^{\nu}(a_{B}\boldsymbol{k_1})$. The corresponding Fourier transform with respect to $\Delta\boldsymbol{q}$ reads:
\begin{widetext}
    \begin{equation}
\label{eq:fourier_transform_electron_hole_wavefunctions}
\begin{aligned}
    &\frac{A}{4\pi^2}\int d^2k_1 \varphi^{*\,\nu}(a_{B}\boldsymbol{k_1})\left[\frac{1}{4\pi^2}\int d^2 \Delta q e^{i\Delta \boldsymbol{q}\cdot \boldsymbol{r}}\varphi^{\nu'}\!\left(a_{B}\boldsymbol{k_1}-a_{h}\Delta\boldsymbol{q}\right) - \frac{1}{4\pi^2}\!\int \! d^2 \Delta q e^{i\Delta \boldsymbol{q}\cdot \boldsymbol{r}}\varphi^{\nu'}\!\left(a_{B}\boldsymbol{k_1}+ a_{el} \Delta\boldsymbol{q}\right)\right]
    \\
    =&A\left[\frac{\varphi^{*\,\nu}\!\left(\frac{\boldsymbol{r}}{a_{h}}\right)\varphi^{\nu'}\!\left(-\frac{\boldsymbol{r}}{a_{h}}\right)}{a_{h}^2} - \frac{\varphi^{*\,\nu}\!\left(-\frac{\boldsymbol{r}}{a_{el}}\right)\varphi^{\nu'}\!\left(\frac{\boldsymbol{r}}{a_{el}}\right)}{a_{el}^2}\right]
    \end{aligned}
\end{equation}
\end{widetext}
where we used Fubini's theorem \cite{billingsley_probability_1995} and the scaling and shifting property of the Fourier transform\cite{bronstejn_taschenbuch_2020}.
The contributions in Eq. \ref{eq:fourier_transform_electron_hole_wavefunctions} scale with the effective electron/hole Bohr radius respectively.\\
Restricting to the diagonal s-like contributions, where $\varphi^{\nu}(-\boldsymbol{r}) = \varphi^{\nu}(\boldsymbol{r})$, we find the internal charge distribution of the exciton built up from electron and hole charge distribution:
    \begin{equation}
\label{eq:fourier_transform_to_total_charge_distribution}
\begin{aligned}
&\mathcal{F}\!\left[\sum_{\boldsymbol{k_1}}\bigl(\varphi^{\nu}_{\boldsymbol{k_1}}\! \varphi^{*\,\nu}_{ \boldsymbol{k_1}+\alpha\Delta\boldsymbol{q}}\!\! -\varphi^{*\,\nu}_{  \boldsymbol{k_1}}\! \varphi^{\nu}_{\boldsymbol{k_1}+ \beta \Delta\boldsymbol{q}}\bigr) \right]\!(\boldsymbol{r}) 
\\
=& A \left(\rho^{\nu}_{h} (\boldsymbol{r})\!- \!\rho^{\nu}_{el}(\boldsymbol{r})\right) 
    \end{aligned}
\end{equation}
with $\rho^{\nu}_{h} (\boldsymbol{r}) = \frac{\left|\varphi^{\nu}\left(\frac{\boldsymbol{r}}{a_{h}}\right)\right|^2}{a_{h}^2} $ and  $ \rho^{\nu}_{el}(\boldsymbol{r}) = \frac{\left|\varphi^{\nu}\left(\frac{\boldsymbol{r}}{a_{el}}\right)\right|^2}{a_{el}^2}$.
Obviously, this expression is proportional to the total charge distribution $\rho^{\nu}_{tot}(\boldsymbol{r}) = \rho^{\nu}_{h} (\boldsymbol{r})- \rho^{\nu}_{el}(\boldsymbol{r})$.
However, in contrast to Eq. \ref{eq:fourier_transform_to_total_charge_distribution}, in the spectrum, Eq. \ref{eq:pumped_spectrum_final}, the absolute square of the convolution appears and therefore we will get access to the autocorrelation of the charge distribution:
    \begin{equation}
\label{eq:autocorrelation_charge_dist}
\begin{aligned}
&\mathcal{F}\!\left[\left|\sum_{\boldsymbol{k_1}}\bigl( \varphi^{*\,\nu}_{ \boldsymbol{k_1}+\alpha\Delta\boldsymbol{q}}\varphi^{\nu}_{\boldsymbol{k_1}}\!\! -\!\varphi^{*\,\nu}_{  \boldsymbol{k_1}} \varphi^{\nu}_{\boldsymbol{k_1}+ \beta \Delta\boldsymbol{q}}\bigr) \right|^2\right]\!(\boldsymbol{r}) 
\\
=&\frac{A^2}{4\pi^2}\!\int\! d^2\!\Delta q \left|\int d^2r' \rho^{\nu}_{tot}(\boldsymbol{r'}) e^{\!-i\Delta\boldsymbol{q}\cdot\boldsymbol{r'}} \right|^2\!\!e^{i\Delta\boldsymbol{q}\cdot\boldsymbol{r}}
\\
=&\left(NA_{UC}\right)^2\int d^2r' \rho^{\nu}_{tot}(\boldsymbol{r'})  \rho^{\nu}_{tot}(\boldsymbol{r'}-\boldsymbol{r}) 
    \end{aligned}
\end{equation}
scaled by the factor $A^2=\left(NA_{UC}\right)^2$.
The diagonal contribution of the spectrum, $\nu=\nu'$ in Eq. \ref{eq:pumped_spectrum_final}, can be expressed in a comparable form to a space-time correlation function\cite{schulke_electron_2007}:
\begin{widetext}
    \begin{equation}
\label{eq:pumped_spectrum_final_alternate_form_correlation_function}
\begin{aligned}
S_{\text{pump}}(\Delta\boldsymbol{q},\omega)
=&\left|\boldsymbol{A}^{\! \perp}_{\! X} \omega_{X}\! \frac{ r_{0}}{ R }N\right|^2  \!\frac{1}{2\pi}\! \int\! d\tau e^{i\Delta\omega \tau} \sum_{\nu} \braket{P^{\dagger\, \nu}_{\boldsymbol{0}}(\tau) P^{\nu}_{\boldsymbol{0}}(0)} e^{\!-i\Omega^{\nu}_{\Delta\boldsymbol{q}}\tau}\! \int\! d^2\Delta q e^{\!-i\Delta\boldsymbol{q}\cdot\boldsymbol{r}} G^{\nu}(\boldsymbol{r})
    \\
    \end{aligned}
\end{equation}
\end{widetext}
where
\begin{equation}
    G^{\nu}(\boldsymbol{r}) = A_{UC}^2 \int d^2r'  \rho^{\nu}_{tot}(\boldsymbol{r'})  \rho^{\nu}_{tot}(\boldsymbol{r'}-\boldsymbol{r}) 
\end{equation}
inherits the autocorrelation of the total excitonic charge distribution of the excitonic state $\nu$. To arrive at Eq. \ref{eq:pumped_spectrum_final_alternate_form_correlation_function} we again used Eq. \ref{eq:limit_lorentzian_to_fourier_space}. \\
We now use Eq. \ref{eq:pumped_spectrum_final_alternate_form_correlation_function} to analyze to which extent the internal charge distribution of an exciton is encoded in the x-ray scattering spectra:
First, as a toy example, we restrict the evaluation to contributions from 1s-excitons thereby neglecting all interference contributions from other excitons.   

\begin{figure}[ht!]
    \centering
    \includegraphics[width=\linewidth]{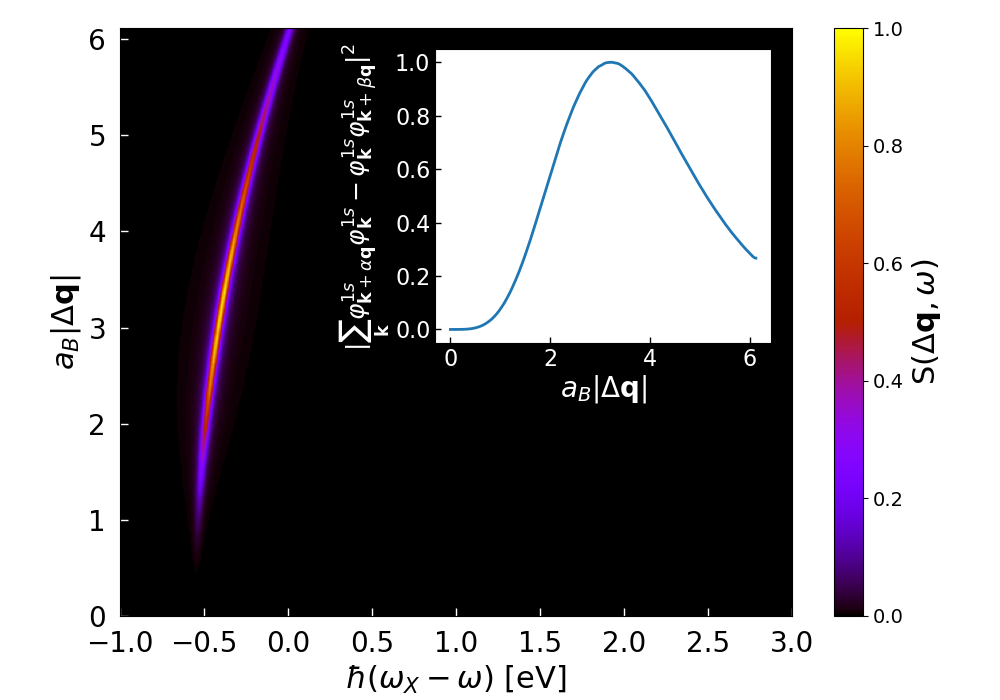}
    \caption{Differential scattering spectrum of WS$_2$ with contributions from 1s-excitons only. The spectrum is normalized with respect to the maximum value. In the inset we see the absolute square of the convolution of the 1s-exciton wavefunctions, which could be detected by integrating the spectrum with respect to the energy $\hbar\omega$.}
  \label{fig:spectrum_pump_for_WS2_only_1s}
\end{figure}
Fig. \ref{fig:spectrum_pump_for_WS2_only_1s} shows the idealized situation of the differential scattering spectrum where exclusively contributions of the 1s-exciton are included. 
Integration over the energy of the $\nu=1s$ peak in Eq. \ref{eq:pumped_spectrum_final}, gives access to the absolute square of the convolution of the 1s-excitonic wave functions $\left|\sum_{\boldsymbol{k_1}}\bigl( \varphi^{*\,1s}_{ \boldsymbol{k_1}+\alpha\Delta\boldsymbol{q}}\varphi^{1s}_{\boldsymbol{k_1}}\! -\varphi^{*\,1s}_{  \boldsymbol{k_1}} \varphi^{1s}_{\boldsymbol{k_1}+ \beta \Delta\boldsymbol{q}}\bigr) \right|^2$, which is shown in the inset in Fig. \ref{fig:spectrum_pump_for_WS2_only_1s}. 
A Fourier transform of the inset in Fig. \ref{fig:spectrum_pump_for_WS2_only_1s} to real space results in the autocorrelation function of the total excitonic charge distribution. To study which experimental features observed in the scattering can be related to excitonic properties, Fig. \ref{fig:1s_exciton_Fourier_trafo_of_convo} shows a comparison of this autocorrelation, inset Fig \ref{fig:spectrum_pump_for_WS2_only_1s}, to the total charge distribution of the 1s exciton: 

\begin{figure}[ht!]
    \centering
    \includegraphics[width=\linewidth]{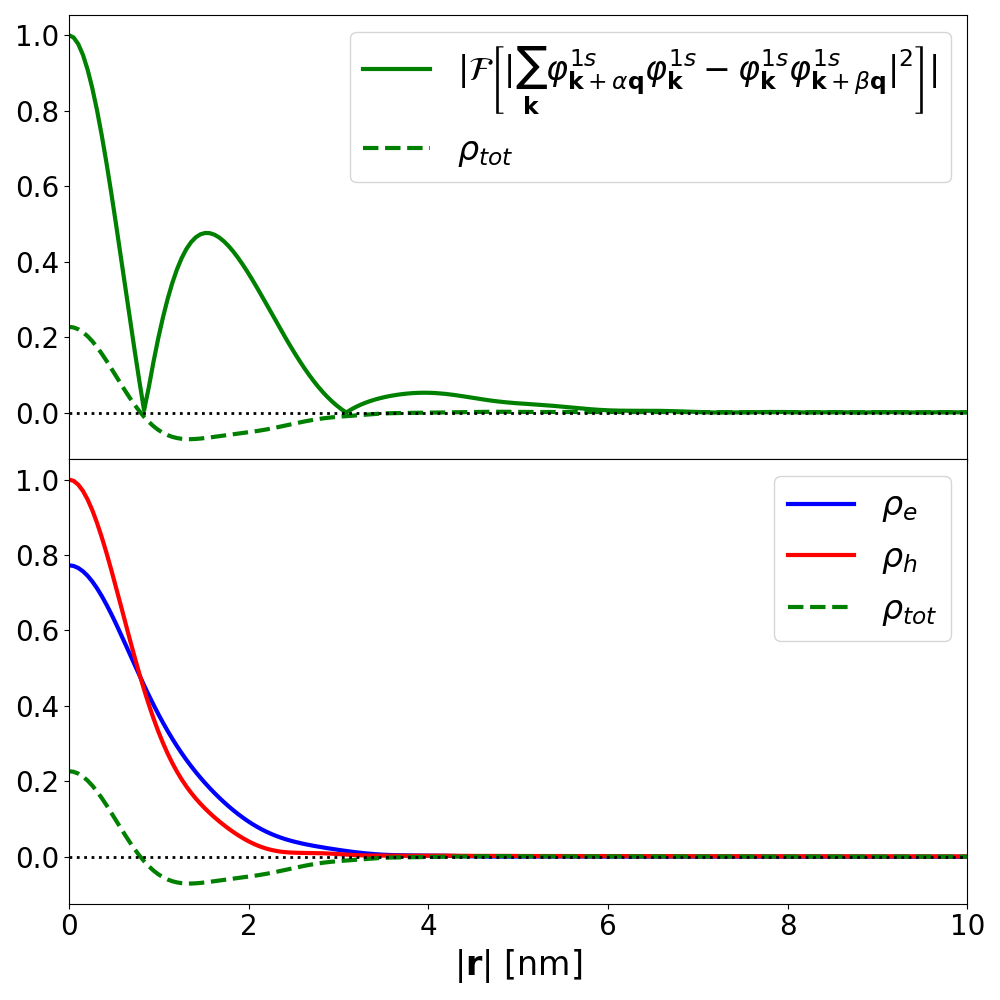}
    \caption{\textbf{Top:} Magnitude of the autocorrelation of the total charge distribution of the 1s-exciton (solid green) compared to the numerically calculated total charge distribution (dashed green). \textbf{Bottom}: Comparison of the charge distribution of electron (blue), hole (red) and the total charge distribution (dashed green). }
  \label{fig:1s_exciton_Fourier_trafo_of_convo}
\end{figure}
The top panel in Fig. \ref{fig:1s_exciton_Fourier_trafo_of_convo} shows both, the numerically calculated internal charge distribution of the 1s exciton (dashed green) in units of the elementary charge and the corresponding magnitude of the autocorrelation (solid green) normalized to 1. For comparison, in the bottom panel the pure electronic charge distribution (blue) and the hole charge distribution (red) are shown adding to the total charge distribution (dashed green). The spatial imbalance between electron and hole distributions is observable, whereas the hole distribution shows a stronger localization in real space. The comparison in the top panel enables the attribution of features of the autocorrelation of the charge distribution to the charge distribution itself: The first local minimum, reached at approximately $\SI{0.85}{\nm}$, is closely associated to the point where the charge distribution changes its sign and corresponds to the position of an equal electron and hole density, cp. bottom panel of Fig. \ref{fig:1s_exciton_Fourier_trafo_of_convo}.
The point, where the autocorrelation finally reaches 0 and does not change anymore, corresponds to the diameter of the total charge distribution of the exciton.
\\
Clearly, the discussed restriction to 1s excitons is an idealized situation because in our evaluation, the signal is only detected in the energy range of the 1s-exciton. In reality, also interferences with other excitons occur depending on the pumping energy. More details are now discussed in the following section \ref{sec:different_pump_energies}

\subsection{Variation of the pump energy}
\label{sec:different_pump_energies}
\begin{figure}[!ht]
    \centering
    \begin{subfigure}[b]{0.49\textwidth}
            \centering
    \includegraphics[width=0.99\linewidth]{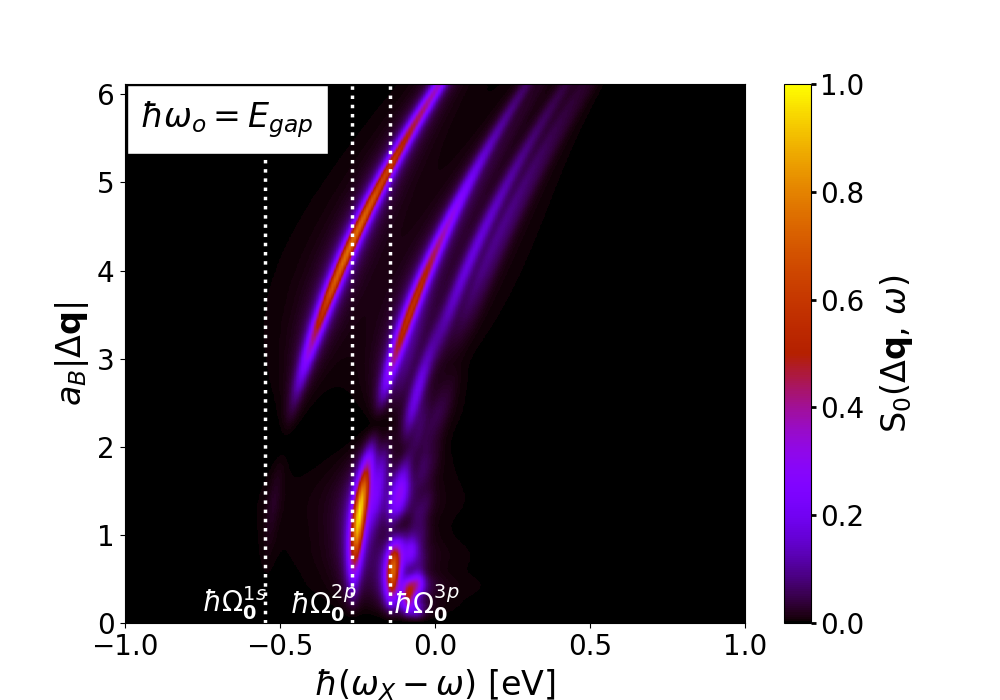}
            \caption{}
    \end{subfigure} 
            \begin{subfigure}[b]{0.49\textwidth}
            \centering
        \includegraphics[width=0.99\linewidth]{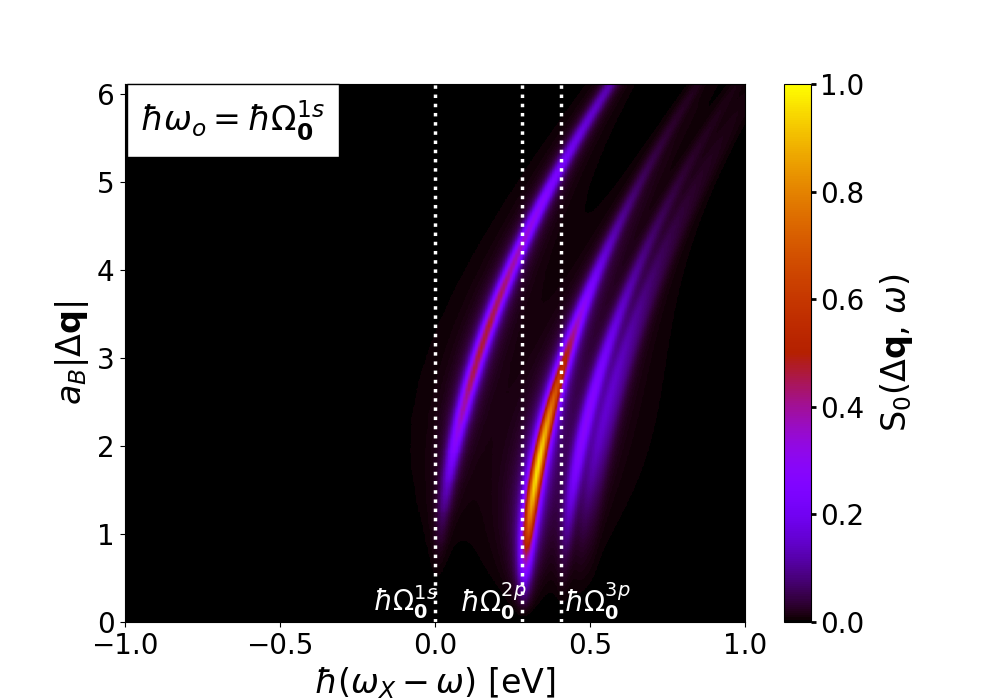}
            \caption{}
    \end{subfigure}
    \caption{$S_{\text{pump}}(\Delta\boldsymbol{q},\omega)$ for different pumping energies: \textbf{(a)} The pump energy corresponds to the band gap energy $\hbar\omega_{o}=E_{\text{gap}}$. \textbf{(b)} The pump energy corresponds the binding energy of the 1s-exciton $\hbar\omega_{o}=\hbar\Omega^{1s}_{\boldsymbol{0}}$. The white dashed lines correspond to the position of the binding energies of different excitonic states. For reasons of visibility the contribution from the 1s-exciton was scaled by a factor of 10 in both spectra.}
    \label{fig:pumping_with_egap_1s_2s}
\end{figure}

In the following, we discuss the spectra $S_{\text{pump}}(\Delta\boldsymbol{q},\omega)$, Eq. \ref{eq:pumped_spectrum_final} involving all possible excitonic contributions, for different pump energies $\hbar\omega_{o}$ at the band edge $\hbar\omega_{o}=E_{gap}$ and at the 1s-exciton resonance $\hbar\omega_{o}=\hbar\Omega^{1s}_{\boldsymbol{0}}$, cp. Fig. \ref{fig:pumping_with_egap_1s_2s}a  and Fig \ref{fig:pumping_with_egap_1s_2s}b respectively.

Fig. \ref{fig:pumping_with_egap_1s_2s} shows a comparison of two spectra for different pump energies $\hbar\omega_{o}$. As we are interested in studying the charge distribution of the 1s-exciton, we scaled this contribution (leftmost contribution) in both spectra for the sake of visibility by a factor of 10, cp.  Fig. \ref{fig:pumping_with_egap_1s_2s}a to Fig. \ref{fig:spectrum_pump_for_WS2}. For a direct comparison we show in Fig. \ref{fig:pumping_with_egap_1s_2s}a contributions of the pumped spectrum already obtained in the previous section, where $\hbar\omega_{o}=E_{gap}$. Fig. \ref{fig:pumping_with_egap_1s_2s}b shows the contributions of the pumped spectrum when pumping resonantly the 1s exciton, $\hbar\omega_{o} = \hbar \Omega^{1s}_{\boldsymbol{0}}$. The white dashed lines in Fig. \ref{fig:pumping_with_egap_1s_2s} correspond to different exciton binding energies and their energetical position for different pump energies.
The overall appearance of the spectrum changes drastically when changing the pump energy $\hbar\omega_{o}$: We observe a shift of the spectral weight with pump frequency $\omega_{o}$ in the energy $\hbar(\omega_{X}\! -\! \omega)$ and momentum transfer $\Delta \boldsymbol{q}$. In general, each contribution to the spectra in Fig. \ref{fig:pumping_with_egap_1s_2s} correspond to a convolution between the optically pumped exciton, which is always an s-like exciton, and all other excitons, i.e. including also optically forbidden excitons, cp. Eq. \ref{eq:pumped_spectrum_final}. When pumping the 1s-exciton, cp. Fig. \ref{fig:pumping_with_egap_1s_2s}b, each observed feature is a convolution of the 1s-exciton with all other excitons, labeled by state $\nu'$ in Eq. \ref{eq:pumped_spectrum_final} including also excitons with vanishing optical oscillator strength and unpumped excitons with finite oscillator strength. Unpumped states $\nu'$ are energetically distinguishable in the spectrum due to their individual binding energies, cp. dashed lines, whereas the contributions from the pumped exciton are found at vanishing energy transfer $\hbar(\omega_{X}\! -\! \omega)$. The strongest contributions in the spectrum in Fig. \ref{fig:pumping_with_egap_1s_2s} result from p-like excitons dominating the spectrum by their interference with the pumped s-like excitons. 
This can be understood by inspecting  the convolution of the excitonic wavefunctions: $    \sum_{\boldsymbol{k_1}}\varphi^{*\,\nu}_{ \boldsymbol{k_1}} \left(\varphi^{\nu'}_{\boldsymbol{k_1}-\alpha\Delta\boldsymbol{q}} - \varphi^{\nu'}_{\boldsymbol{k_1}+ \beta \Delta\boldsymbol{q}}\right)$. The term in the parentheses vanishes for vanishing momentum transfer $\Delta\boldsymbol{q}$ and determines the behaviour of the spectral contributions: While s-like wavefunctions have an even symmetry $\varphi^{s}(-\boldsymbol{k)} = \varphi^{s}(\boldsymbol{k)}$, p-like wavefunctions have an odd symmetry $\varphi^{p}(-\boldsymbol{k)} = - \varphi^{p}(\boldsymbol{k)}$. Even though, $\alpha$ and $\beta$ are not equal, they are of the same order. 
Lets assume $\alpha=\beta$ and discuss $\boldsymbol{k_1}=\boldsymbol{0}$, where we find:
$\left(\varphi^{p}_{-\alpha\Delta\boldsymbol{q}}\! - \varphi^{p}_{ \alpha \Delta\boldsymbol{q}}\right) = 2\varphi^{p}_{-\alpha\Delta\boldsymbol{q}} $. This explains the dominance of the convolution with p-like excitons. A similar argument for s-like excitons yields a vanishing contribution.\\
Nevertheless, due to the possibility of adjusting the pumping energy and detecting at a specific energy, experiments are able to get access to selected excitonic states, including charge distributions. In the following, we want to
discuss a possibility to detect the 1s charge distribution under the interference of all other excitons. As the contributions of s- and p-excitons are found at different energies, cp. Fig. \ref{fig:spectrum_pump_for_WS2_only_1s} and Fig. \ref{fig:pumping_with_egap_1s_2s}b, we will use an energy integrated signal $S^{1s}_{\text{pump}}(\Delta\boldsymbol{q}) = \int_{(1s)} d\omega S_{\text{pump}}(\Delta\boldsymbol{q},\omega) $ where we collect only contributions close to the 1s-exciton resonance. The results shown are thus only dependent on the momentum transfer:

\begin{figure}[th!]
    \centering
    \includegraphics[width=0.99\linewidth]{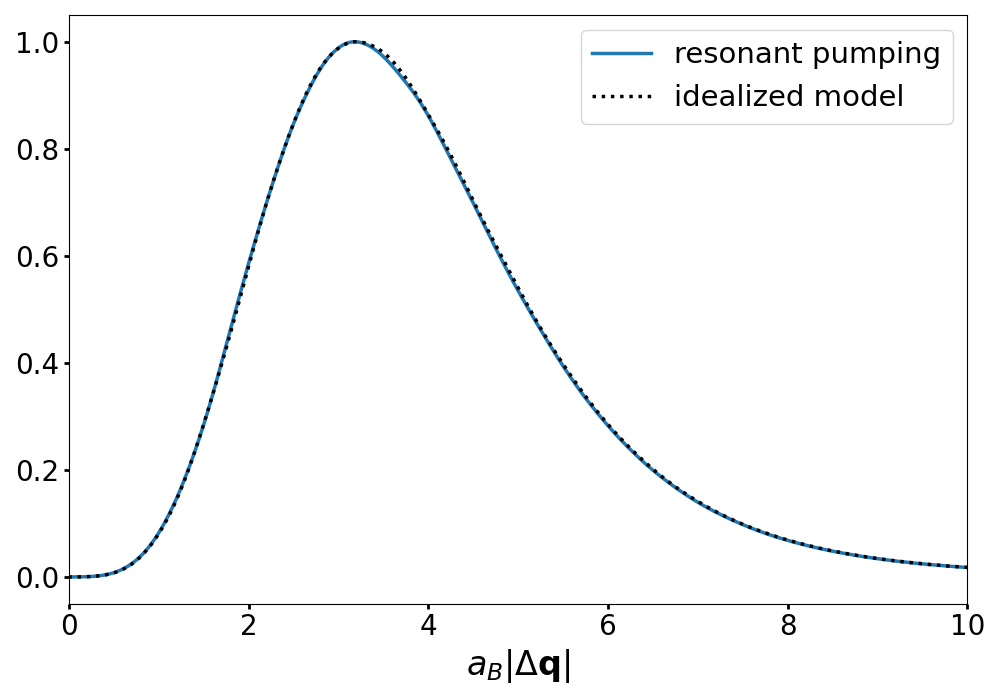}
    \caption{Comparison of the momentum transfer dependent spectrum $S^{1s}_{\text{pump}}(\Delta\boldsymbol{q})$ when pumping resonantly the 1s-exciton $\omega_{o}=\Omega^{1s}_{\boldsymbol{0}}$ (blue) and the spectral contribution when including only the 1s-exciton (black dotted). Both contributions are nearly perfectly matching.}
    \label{fig:comparison_only_1s_to_pumping_1s}
\end{figure}
Fig. \ref{fig:comparison_only_1s_to_pumping_1s} shows the comparison of the momentum transfer dependent spectrum $S^{1s}_{\text{pump}}(\Delta\boldsymbol{q})$ for pumping the 1s-exciton resonantly and detecting close to the resonance (blue) and the idealized situation where only 1s-excitons are included in Eq. \ref{eq:pumped_spectrum_final} (black dotted). The latter was discussed in section \ref{sec:charge_distribution_of_exciton} and corresponds to the spectral features, which can be directly connected to the charge distribution of the exciton. As both contributions are almost identical,
we obviously also get access to the spectral contributions of the charge distribution when pumping the exciton resonantly, cp. section \ref{sec:charge_distribution_of_exciton}.\\
This can be even better evaluated by looking at the Fourier transform to real space:
\begin{figure}[th!]
    \centering
    \includegraphics[width=0.99\linewidth]{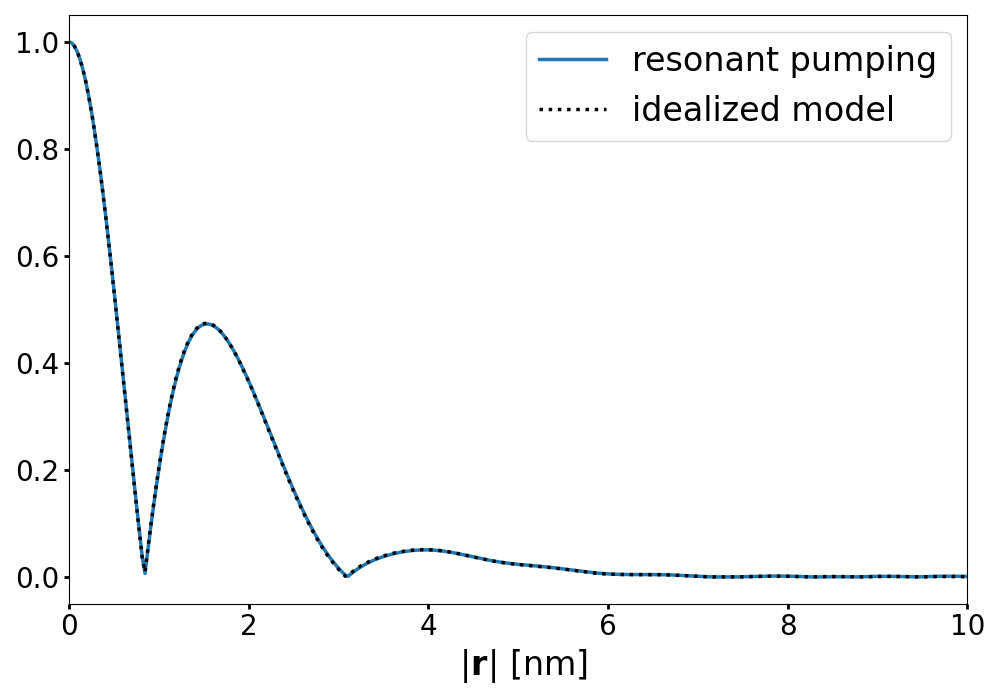}
    \caption{Fourier transforma of Fig. \ref{fig:comparison_only_1s_to_pumping_1s} to real space for the comparison of resonantly pumping the 1s-exciton $\omega_{o}=\Omega^{1s}_{\boldsymbol{0}}$ (blue) with the idealized model where only the 1s-exciton contributions are included (black dotted).}
    \label{fig:comparison_only_1s_to_pumping_1s_fourier_trafo}
\end{figure}
Fig. \ref{fig:comparison_only_1s_to_pumping_1s_fourier_trafo} shows the Fourier transform of both contributions of Fig. \ref{fig:comparison_only_1s_to_pumping_1s} to real space. It is easy to verify that both results match perfectly. Consequently, by pumping the exciton resonant while detecting close to its energy, it is possible to utilize this result for the interpretation of the associated excitonic charge distribution.

\section{Conclusion}
We studied the scattering of x-rays from optically pumped excitons for the example of atomically thin WS$_2$. In addition to the intrinsic scattering spectrum, dominated by valence band electrons and scattering from unpumped exciton resonances we also find quasi-elastic contributions occuring at low energy transfer, that can be attributed directly to the optical pumping of excitons. These new contributions have a direct connection to the Fourier transform of the spatial charge distribution formed by electrons and holes within the exciton. Interferences between excitonic wave functions of different states, in particular p-excitons, contribute to the spectrum but are energetically distinguishable from the contributions of the charge distribution.  
In contrast to time-resolved resonant inelastic x-ray scattering \cite{beye_time_2013,chen_theory_2019}, there is little work on IXS of optically pumped systems far from resonances. It could be beneficial to examine such systems as one could extent these studies to interacting quasiparticles at high densities.
\section{Acknoledgments}
J. S. and A. K. acknowledge financial support from the Deutsche Forschungsgemeinschaft (DFG) through Project No. 527838492. A. K. was supported by the ’Stephenson Distinguished Visitor Programme’ at DESY in Hamburg (Germany).

\bibliographystyle{ieeetr}
\bibliography{references2}

\appendix
\section{Fourier transform and excitonic wavefunctions}
\label{app:fourier_transform_and_Excitons}
In this contribution we make use of the discrete and continuous Fourier transform.
\\For the time- and frequency domain we use the following definition:
\begin{equation}
    f(t) = \frac{1}{2\pi}\int d\omega f(\omega) e^{\!-i\omega t}, \qquad f(\omega) = \int dt f(t) e^{i\omega t}
\end{equation}
In the real- and momentum space we use for the continuous Fourier transform:
\begin{equation}
  f(\boldsymbol{r}) = \frac{1}{(2\pi)^n}\int d^n k\, f_{\boldsymbol{k}} e^{i\boldsymbol{k}\cdot \boldsymbol{r}},  \qquad f_{\boldsymbol{k}} = \int d^{n}r f(\boldsymbol{r}) e^{\!-i\boldsymbol{k}\cdot \boldsymbol{r}}
\end{equation}
where $n$ is the dimensi
on of the real- or momentum space. \\
For the discrete Fourier transform we use:
\begin{equation}
\label{eq:Fourier_transform_discrete}
  f(\boldsymbol{r}) = \frac{1}{\mathcal{V}^{(n)}} \sum_{\boldsymbol{k}} f_{\boldsymbol{k}} e^{i\boldsymbol{k}\cdot \boldsymbol{r}},  \qquad f_{\boldsymbol{k}} = \frac{\mathcal{V}^{(n)}}{N} \sum_{\boldsymbol{r}} f(\boldsymbol{r}) e^{\!-i\boldsymbol{k}\cdot \boldsymbol{r}}
\end{equation}
where $\mathcal{V}^{(n)}$ is the "volume" of the whole space with respect to dimension $n$, so e.g. $\mathcal{V}^{(2)} = A$.
\\
In this contribution, we use unitless excitonic wavefunctions in reciprocal space $\varphi^{\nu}_{\boldsymbol{\kappa}}$ with the following normalization:
\begin{equation}
\label{eq:app_normalization_exciton_wavefunction}
    \begin{aligned}
        &\sum_{\boldsymbol{\kappa}} \varphi^{*\, \nu}_{\boldsymbol{\kappa}} \varphi^{\mu}_{\boldsymbol{\kappa}} \hspace{0.12cm}= \delta^{\nu,\mu}
        \\
        &\sum_{\nu} \varphi^{*\, \nu}_{\boldsymbol{\kappa}_1} \varphi^{\nu}_{\boldsymbol{\kappa}_2} = \delta_{\boldsymbol{\kappa}_1,\boldsymbol{\kappa}_2}
    \end{aligned}
\end{equation}
By using Eq. \ref{eq:Fourier_transform_discrete} for the Fourier transform of exciton wavefunctions:
\begin{equation}
    \sum_{\boldsymbol{\kappa}}\varphi^{\nu}_{\boldsymbol{\kappa}} e^{i\boldsymbol{\kappa}\cdot\boldsymbol{r}} = A \varphi^{\nu}(\boldsymbol{r})
\end{equation}
it gets obvious that in real space $\varphi^{\nu}(\boldsymbol{r})$ has to have units of an inverse area, so $\si{m^{-2}}$. In fact, the exciton wavefunctions are in reciprocal space not only dependent on $\kappa$ but on $a_{B}\kappa$, where $a_{B}$ is the Bohr radius of the exciton. Therefore, when using the scaling property of the Fourier transform we know that we can write $\varphi^{\nu}(\boldsymbol{r}) = \frac{\tilde{\varphi}^{\nu}(\boldsymbol{r})}{a_{B}^2}$, where $\tilde{\varphi}^{\nu}(\boldsymbol{r})$ is again a unitless function. This can be verified by using the scaling property of the Fourier transform:
\begin{equation}
\label{eq:app:fourier_transform_exciton_wavefunction_bohr_radius}
     \sum_{\boldsymbol{\kappa}}\varphi^{\nu}_{a_{B}\boldsymbol{\kappa}} e^{i\boldsymbol{\kappa}\cdot\boldsymbol{r}} = \frac{A}{a_{B}^2} \tilde{\varphi}^{\nu}\!\left(\frac{\boldsymbol{r}}{a_{B}}\right)
\end{equation}

\section{Free electron gas limit}
\label{app:electron_gas_limit}
Using unperturbed Bloch electrons of a single band: $ a^{\dagger\lambda}_{\boldsymbol{k}}(t) \to a^{\dagger}_{\boldsymbol{k}}(t) = a^{\dagger}_{\boldsymbol{k}} e^{i\omega_{\boldsymbol{k}}t}$ where we introduced $\omega_{\boldsymbol{k}} = \epsilon_{\boldsymbol{k}}/ \hbar$ for the free electron operators $a^{(\dagger)}_{\boldsymbol{k}}$, a Hartree-Fock approximation in Eq. \ref{eq:definition_of_G_current_density_correlation} provides
\begin{equation}
\begin{aligned}
        &\braket{ a^{\dagger}_{\boldsymbol{k_1}}(t\!+\! \tau) a_{\boldsymbol{k_1} +\Delta \boldsymbol{q}}(t\!+\! \tau) a^{\dagger}_{\boldsymbol{k_2} +\Delta \boldsymbol{q} }(t)a_{\boldsymbol{k_2}}(t) }  
        \\
    \approx& f_{\boldsymbol{k_1} } f_{\boldsymbol{k_2}} \delta_{\Delta\boldsymbol{q},\boldsymbol{0}} +\delta_{\boldsymbol{k_1},\boldsymbol{k_2}} f_{\boldsymbol{k_1}} (1-f_{\boldsymbol{k_1} +\Delta \boldsymbol{q}} ) e^{i(\omega_{\boldsymbol{k_1}}-\omega_{\boldsymbol{k_1}+\Delta\boldsymbol{q}})\tau} 
        \end{aligned}
\end{equation}
where we introduced a spatial homogeneity by $\braket{ a^{\dagger}_{\boldsymbol{k_1}}   a_{\boldsymbol{k_2}} }\approx f_{\boldsymbol{k_1}} \delta_{\boldsymbol{k_1},\boldsymbol{k_2}}$ and $f_{\boldsymbol{k_1}}$ is the electronic Fermi-function.\\

Introducing $N_{0} = \sum_{\boldsymbol{k}} f_{\boldsymbol{k}}$ as the number of electrons and $  f_{\boldsymbol{k_1}} (1-f_{\boldsymbol{k_1}+\Delta\boldsymbol{q}})  = g(\Delta \omega) (f_{\boldsymbol{k_1}}-f_{\boldsymbol{k_1}+\Delta \boldsymbol{q }})$ where $g(\Delta \omega) = 1/(1-e^{\!-\hbar \Delta \omega/k_BT})$ and $ \Delta \omega =\omega_{X} - \omega$, we find:
\begin{widetext}
    \begin{equation}
\label{eq:spectrum_free_electron_gas_limit}
\begin{aligned}
      S(\Delta\boldsymbol{q}, \omega)  =&2\pi\left|\boldsymbol{A}^{\! \perp}_{\! X} \omega_{X}\! \frac{ r_{0}}{ R }\right|^2 \delta_{\Delta\boldsymbol{q},0}  N_{0}^2  \delta(\omega-\omega_{X})
      \\
      +&2\pi\left|\boldsymbol{A}^{\! \perp}_{\! X} \omega_{X}\! \frac{ r_{0}}{ R }\right|^2 g(\Delta \omega)\sum_{\substack{ \boldsymbol{k_1} }} (f_{\boldsymbol{k_1}}-f_{\boldsymbol{k_1}+\Delta \boldsymbol{q }}) \bigl|\braket{\boldsymbol{k_1}| \boldsymbol{k_1}\! + \! \Delta \boldsymbol{q}} \bigr|^2  \delta(\omega-\omega_{X}-\omega_{\boldsymbol{k_1}}+\omega_{\boldsymbol{k_1}+\Delta\boldsymbol{q}})
\end{aligned}
\end{equation}
\end{widetext}
This limit contains the well known scattering signal from a free electron gas where the first term accounts for elastic scattering with energy and momentum conservation of the incoming and outgoing wave. Non-elastic scattering in the second term allows for momentum- and energy transfer. Eq. \ref{eq:spectrum_free_electron_gas_limit} is in agreement with the well known results of Refs. \cite{cardona_light_1983,sturm_dynamic_1993,schulke_electron_2007}.

\section{Material parameters}
\label{app:material_parameters}
In the following all parameters are shown which have been used for the calculation of the Wannier equation (Eq. \ref{eq:wannier_equation}) and the unpumped and pumped spectra. \\
\begin{table}[!ht]
\begin{tabular}{||c | c ||} 
 \hline
  & WS$_2$ \\ 
 \hline\hline
 m$_e$/m$_{0}$\cite{kormanyos_k_2015} & 0.26 \\ 
 \hline
  m$_h$/m$_0$\cite{kormanyos_k_2015}& 0.35  \\ 
   \hline
  $\epsilon_{\text{TMDC}}$\cite{kumar_tunable_2012}& 11.5  \\ 
     \hline
  $h_{\text{TMDC}}$\cite{rasmussen_computational_2015}& $\SI{0.63}{\nm}$  \\ 
 \hline
   $\hbar \gamma \,(\SI{300}{\K})$\cite{selig_excitonic_2016}   & $\SI{23.2}{meV}$ \\ 
 \hline
    $\Delta\epsilon^{1s}$  & $\SI{550}{meV}$ \\ 
 \hline
     $\Delta\epsilon^{2s}$  & $\SI{222}{meV}$ \\ 
 \hline
     $\Delta\epsilon^{3s}$  & $\SI{125}{meV}$ \\ 
 \hline
     $\Delta\epsilon^{4s}$  & $\SI{81}{meV}$ \\ 
 \hline
     $\Delta\epsilon^{5s}$  & $\SI{57}{meV}$ \\ 
 \hline
     $\Delta\epsilon^{2p}$  & $\SI{269}{meV}$ \\ 
 \hline
     $\Delta\epsilon^{3p}$  & $\SI{144}{meV}$ \\ 
 \hline
     $\Delta\epsilon^{4p}$  & $\SI{90}{meV}$ \\ 
 \hline
     $\Delta\epsilon^{5p}$  & $\SI{62}{meV}$ \\ 
 \hline
     $\Delta\epsilon^{3d}$  & $\SI{163}{meV}$ \\ 
 \hline
\end{tabular}
 \caption{Values of the effective mass of electron/hole (m$_e$/m$_h$), the dielectric constant of the TMDC's, the height of the TMDC and the linewidth which are used for the calculation of the spectra. Below that we show the numerical values of the binding energies of the different excitonic states which are used within this contribution.}
\label{tab:values_of_effective_masses_and_binding_energies}
\end{table}
\noindent
In this contribution we assumed that the linewidth does not depend on the exciton quantum number $\nu$ or wavevector $\boldsymbol{Q}$ and assumed $\hbar \gamma^{\nu}_{\boldsymbol{Q}} \approx \hbar \gamma$.

\end{document}